\documentclass[12pt]{article}
\usepackage{graphicx}
\input epsf
\def\c2m{cm$^{2}$}
\def\cm2{cm$^{2}$}
\def\cm3{cm$^{3}$}

\def\26Al{$^{26}$Al}

\begin{document}
\baselineskip 0.65cm

\begin{center}
{\Large  
  {\bf The Earth-Moon system during the Late Heavy Bombardment period }} \\
{-- geochemical support for impacts dominated by comets.}\\
\end{center}
\vspace{0.5cm}
\noindent
{\bf Uffe Gr{\aa}e J{\o}rgensen$^{1,*}$, Peter W.U.\ Appel$^2$, 
Yuichi Hatsukawa$^3$,  Robert Frei$^4$,
Masumi Oshima$^3$, Yosuke Toh$^3$, Atsushi Kimura$^3$ } \\

\vspace{0.5cm}

\noindent
{\small
$^1${\em Niels Bohr Institute and Centre for Star and Planet Formation, 
 Juliane Maries Vej 30, 2100 Copenhagen, Denmark},\\
$^*$Corresponding author email address: uffegj@nbi.dk\\
$^2${\em Geological Survey of Denmark and Greenland, 
                               {\O}ster Voldgade 10, 1350 Copenhagen, Denmark},\\
$^3${\em Japan Atomic Energy Agency, Tokai, Ibaraki 319-1195, Japan},\\
$^4${\em Institute of Geography and Geology, and Nordic Center for Earth Evolution,
     {\O}ster Voldgade 10, 1350 Copenhagen, Denmark}.\\
 }   
\vspace{1.0cm}

\newpage

\noindent
{\large \bf Abstract.}

The solid planets assembled 4.57 Gyr ago during a period of less 
than 100 Myr, but the bulk of the impact craters we see on the inner 
planets formed much later,
in a narrow time interval between 3.8 and 3.9 Gyr ago, 
during the so-called Late Heavy Bombardment (LHB). 
It is not certain what caused the LHB, and it has not been well
known whether the impactors were comets or asteroids, but our 
present study lend support to the idea that it was comets.
Due to the Earth's higher gravity, the impactors will have hit the 
Earth with $\sim$twice the energy density that they hit the Moon,
and the bombardment will have continued on Earth longer 
than on the Moon.
All solid surface of the Earth will have been completely covered
with craters by the end of the LHB.

However, almost nothing of the Earth's crust from 
even the end of this epoch,
is preserved today. One of the very few remnants, though,
is exposed as the Isua greenstone belt (IGB) and nearby areas in 
Western Greenland. During a field expedition to Isua, we sampled three
types of metasedimentary rocks, deposited $\sim$3.8 billion years ago,
that contain information about the sedimentary river load from larger areas 
of surrounding land surfaces (mica-schist and turbidites) and of 
the contemporaneous seawater (BIF).
Our samples show evidence of the LHB impacts that took
place on Earth, by an average of a seven times enrichment (150 ppt) 
in iridium compared to present day ocean crust (20 ppt).
The clastic sediments show slightly higher enrichment than the chemical 
sediments, which may be due to contamination from admixtures of
mafic (proto-crustal) sources.

We show that this enrichment is in agreement with the lunar cratering
rate and a corresponding extraterrestrial LHB contribution to the 
Earth's Hadean-Eoarchean crust, provided the bulk of the
influx was cometary (i.e., of high velocity and low in CI abundance),
but not if the impactors were meteorites (i.e.\ had velocities and 
abundances similar to present day Earth crossing asteroids).
Our study is a first direct indication of the nature of the LHB
impactors, and the first to find an agreement between the 
LHB lunar cratering rate and the Earth's early geochemical record
(and the corresponding lunar record).
The LHB comets that delivered the iridium we see at Isua will at
the same time have delivered the equivalent of a $\sim$1\,km deep
ocean, and we explain why one should expect a cometary ocean
to become roughly the size of the Earth's present-day ocean, not
only in terms of depth but also in terms of the surface area it covers.
The total impacting mass on the Earth during the LHB will have 
been $\sim$1000\,t/m$^2$.  \\


\newpage

\noindent
{\large \bf Introduction.}

While the lunar craters could in principle represent the end of the 
planetary accretion, most evidence point to the planetary accretion
epoch and the lunar crater formation as being two different 
events in the history of the solar system, separated in time by
several hundred million years. Comparison of the 
relative age and size distribution of craters
throughout the solar system, indicates that the event that created
the ancient lunar craters during the LHB period, (i.e.\ the 
crater-rich highlands and the basins that later became the Mare regions)
also formed the craters on Mars, asteroids, Mercury, and elsewhere
in the solar system; in other words that the LHB was 
unique, heliocentric, and the most violent independent
event that has happened during the whole history of our solar system
since the formation of the planets
(e.g., Gomes et al.\ 2005, Martin et al.\ 2006,
Hartmann et al.\ 2000, Kring \&\ Cohen 2002, Ryder 2002,
Pater \&\ Lissauer 2001).

Very little, however, is known about its effect on the Earth.
Several studies have searched for signs of the LHB on the Earth,
throughout many years, without conclusive evidence and often with 
seemingly contradictory results.
It is often assumed that the traces of the LHB on Earth has been erased
due to Earth's dynamic geology. However, the LHB was so intense, in particular
on the Earth, that it has radically affected the bulk composition of the 
atmosphere and hydrosphere, and probably the crust and mantle, too. 
Judging from the size of the lunar craters,
the largest LHB impacts on Earth will have evaporated the ocean
and stripped off a major fraction of the atmosphere 
(e.g., Zahnle \&\ Sleep 1997, Chyba 1991). There will have been 
$\sim$ 3000 impacts comparable with or larger than the K/T event,
and $\sim$ 1000 tons of cosmic material fell on each m$^2$ of the 
Earth's surface (see calculations below). A few pieces of the Earth's 
present-day crust is formed from material that was part of the 
Earth's solid surface during the end of the LHB period.
The problem of identifying what happened on Earth during the
LHB is therefore not the amount of accessible material, but rather 
difficulties in interpreting the available data, and very uncertain,
or completely lacking, knowledge about the basic astronomical, physical 
and chemical properties of the various possible impactors.

We have collected and analysed three different types of 
sedimentary rock samples from the Earth's oldest
well preserved sedimentary crust; the $\sim$\,3.8\,Gyr old Isua
greenstone belt (IGB) north-west of Nuuk in Greenland.
The sediments were deposited during the end of the LHB period, and
we are here presenting the direct geochemical evidence they show of
what we interpret as a cometary nature of the LHB impactors. \\

\newpage

\noindent
{\large \bf Interpretations of possible atmospheric and crustal signals 
of the LHB.}

The most obvious signal of the LHB, but also one of the most difficult
to interpret, may lie in the Earth's atmosphere, and
we may in fact breathe the gases from the LHB impactors through our 
lungs every day, and we may sip a piece of the impactors every time we
drink a glass of water.
Marty \&\ Meibom (2007) compared the relative 
noble gas abundances in the Earth's atmosphere with that of 
carbonaceous chondrites and comets, and concluded that up to 6\% of the
nitrogen content in our present-day atmosphere may have originated from 
the LHB impactors. They concluded that the bulk of the impactors must 
have been of a composition similar to carbonaceous chondrites, and that 
a mixture with as little as 0.5$\%$ cometary material could explain
the difference between the relative noble gas abundances in the Earth's
mantle and atmosphere. In order to translate these differences 
into percentage of cometary contribution to the LHB, one, however, has
to make assumptions about the relative noble gas abundances interpreted
from a few present-day cometary out-gassing spectra.
This exercise involves estimating the noble gas content, not 
of the possible cometary LHB impactors where they formed (which can
be difficult enough), but of the noble gas abundances in the fractured pieces 
of such comets after they have been disrupted and heated by approaching
the Sun typically several times before impact. On top of this, {\em ``comets"} 
is a concept that hide many different classes of objects, formed from close 
to the orbit of Jupiter, and anywhere out to far beyond the orbit of Neptune,
where the noble gas capture rate, as well as the isotopic ratio of individual
types of molecules, differ widely. The estimates of the LHB cometary
contribution percentage to the present-day composition of the 
atmosphere of the Earth, therefore still include so many 
unknown parameters that the results, in our opinion, mainly should
be understood as an interesting hint that the LHB contribution to the 
Earth's atmosphere may be mainly chondritic, as concluded by Marty \&\
Meibom (2007).
A number of other papers have dealt with the possible cometary contribution
to the origin of the Earth's present atmosphere and hydrosphere in general 
(without specific concern of whether the cometary volatiles were necessarily 
related to the LHB or 
other epochs of the Earth's evolution), but a general consensus is still
lacking (e.g., Pepin 1991, Owen et al.\ 1992, 
Owen \&\ Bar-Nun 2001, Dauphas 2003, Zahnle 2006).
The conclusions are strongly dependent on the detailed assumptions about
the nature of the impacting cometary pieces. 

One could hope that the study of solid rocks would give a more 
straightforward picture than the volatile atmosphere, but also
here the interpretation has been more difficult than anticipated.
Lunar rocks should in principle be the simplest rocks to 
interpret in terms of LHB, because the bulk
of the lunar craters are formed by the LHB impactors and only little 
has changed on the Moon since the end of the LHB. Therefore also the
bulk of the sampled lunar impact melt rocks are due to these impacts, and 
could be expected to be a mixture of lunar crust and representative cosmic
LHB impact material. However, the interpretation of the lunar rock
measurements have been far from simple.
A recent example is the work by Kring \&\ Cohen (2002). By measuring the
abundances of Au, Ge, and Ir in Apollo impact melt samples, they found that
2 of the 6 melts plotted in a 3-element abundance diagram of these elements
fell in the region of iron meteorites, 3 fell in the region of enstatite
meteorites, while one fell outside the region of
any known meteorites. Based on these results, they
concluded that asteroids were the cause of the LHB (as opposed to comets).
In earlier studies (Gros et al.\ 1976, Hertogen et al.\ 1977) 
based on a 3-element abundance 
diagram of other elements, it was concluded that the resemblance was
largest not with iron meteorites as in Kring \&\ Cohen's work,
but with ordinary chondrites (and enstatites) and with C1 carbonaceous 
chondrites, respectively.
However, only relatively little is known about the abundances in the 
comets that can have bombarded the Earth-Moon system during the LHB, 
and even more important is (as we will show below) that
the bulk of impacting comets will leave no chemical
trace on the Moon. This is because the high ratio of cometary impacting 
velocity to lunar escape velocity, will make all the cometary
material being re-ejected into space. The fact that 
some studied melts fall in the same region as some meteorite-types
in a 3-element diagram, therefore give no information about the cometary
contribution.

Alternatively to studying the lunar material from the LHB period, one 
could sample the oldest rocks on Earth, closest in time to the LHB period. 
As opposed to on the Moon, a considerable fraction of the cometary
material will stay on the Earth after impact.
Koeberl et al.\ (1999) looked 
for shock impacted material in the $\sim$3.8 Gyr old 
Isua region and searched for material
with CI-type relative abundances. In spite of the relatively high 
detection limit in their measurements, they did notice excess iridium in
a few of the samples, but because of a failure to find shocked material and 
a relative abundance pattern matching CI composition, they concluded that
the material was not contaminated with LHB impactors, without giving any
alternative explanation of the enhanced iridium. However, neither cometary 
nor asteroid impacts will necessarily result in CI abundance 
patterns (since various types have widely different relative abundances),
and shocked impact material, if identified, is concentrated toward the 
impact site, whereas the non-destructive LHB enrichment of the proto-Isua 
crust is more likely to have been in the form of stratospheric dust fall-out.

A very illustrative example of how unexpected the composition of impacting 
material can be, is the largest known impact ($\sim$100\,m diameter or
$\sim$10$^6$ tons) in historic time (the Tunguska event in 1908).
The most pronounced evidences of its cosmic nature are: a 30 fold increase
(relative Earth's present-day upper crust) in the peat ash iridium
abundance in some of the surrounding swamp peat layers (Korina et al.\
1987, Hou et al.\ 1998), 
a strongly non-terrestrial Pb isotopic pattern (Kolesnikov et al.\ 2005), 
a very peculiar C/Ir ratio (unknown from any other cosmic body but 
interpreted as indicating cometary origin; Rasmussen et al.\ 1999), 
weak CI abundance correlation, and no shocked material.
There is no doubt that the impact was a cosmic event, 
although there are still strong disputes about whether is was 
a carbonaceous-like chondrite or a comet 
(e.g.\ Turco et al 1982, Vasiljev 1998, 
Bronshten 2000, Jopek et al.\ 2008). 
The Tunguska event did result in a local dust
cloud, but did not result in a traceable 
global fall-out (Rasmussen et al.\ 1995), but its mass corresponds to 
only the lower end of the LHB mass distribution. A more typical LHB impactor
(in terms of the total mass contribution) will have corresponded to 
the well studied Cretaceous-Tertiary (K/T) Chicxulub 
impactor that fell, in the region of present-day
Mexico, 65 million years ago. It is likely to have been a cosmic 
body of approximately 10 km in diameter, and 
there will have been several thousands of such impacts on the Earth 
during the LHB. A mixture of material from the Chicxulub
impactor itself and the impacted crust was whirled into the stratosphere,
where it was transported worldwide and over the following several years
fell as a cm-thick layer of dust all over the globe.
The enhanced iridium abundance
in this dust layer was the most important proof that it was due to 
a cosmic impact 
(Alvarez et al.\ 1981, Hansen et al.\ 1988, Frei\& Frei 2002).
There were no associated Pb isotopic anomalies, no reported extreme
C/Ir anomaly, but CI-like PGE abundances. Shocked quartz and feldspar 
have been reported, but their abundances relative to the rest of the 
fall-out material varies with orders of magnitude between the 
different K/T sites at various distances from
the impact location (Bohor et al.\ 1987, Cisowski 1990).
Other studies of early major impacts on Earth include those of
Glikson (2008), Simonson \&\ Glass (2004), and Lowe et al.\ (2003).

We envision that the Earth's atmosphere by the end of the LHB period
must have been rich in cosmic dust that slowly settled over most of the
globe, in analogy to the Tunguska and K/T-events, but not necessarily
identical in composition to any of the two.

Anbar et al.\ (2001) searched for cosmic traces of Ir and Pt in samples
from the Akilia island near Nuuk, approximately 150 km south-west of Isua.
Nutman et al.\ (1996, 1997), and later Mojzsis \&\ Harrison (2000), found the
age of the Akilia suite to be close to 3.85 Gyr, thus being older than the
Isua samples we analyse here and closer in age to the peak of the lunar LHB.
However, this depositional 
age determination was later questioned by Kamber \&\ Moorbath
(1998, 2000) who found an age of 3.65 Gyr. Apart from the 
discussion regarding the depositional age constraints,
also the nature of the rocks was disputed. Fedo \&\ Whitehouse (2002)
found that what was at first interpreted as sediments in ancient seawater
(BIF), were rocks of ultramafic igneous origin that only superficially
resembled BIF. The criticism of the dating and the nature of the rocks
were summarised by Moorbath (2005) together with critical remarks about
Anbar et al.'s interpretation of possible carbon-isotopic traces of life
in the Akilia rocks. Nevertheless, Anbar et al.'s results are
interesting for the present analysis in two respects: 
(1) These old rocks are depleted in iridium relative to the
present-day upper crust, and (2) the lower iridium abundance with a factor
10 or more compared to our Isua measurements is consistent with an
exponentially decaying LHB impact rate only if Kamber \&\ Moorbath's age 
determination of 3.65 Gyr is correct instead of Anbar et al's of 3.85 Gyr.\\

\noindent
{\large \bf W, Hf, and Cr versus iridium.}

Two relatively recent studies searched for signs of the 
LHB in sediment samples from the Isua greenstone 
belt area, with seemingly contradicting results, and 
therefore needs particular attention and comparison with the present study.
Both of the studies aimed at searching for signs of the LHB by 
use of short-lived radiogenic tracers (the
$^{182}$Hf$-^{182}$W system; Schoenberg et al.\ 2002, 
and the $^{53}$Mn$-^{53}$Cr system; Frei \&\ Rosing 2005).

$^{182}$Hf decays to $^{182}$W with a half-life $\tau_{\rm Hf}$ of only 
9\,Myr.  W is siderophile, while Hf is lithophile. After the Earth
melted, W therefore mixed with iron and became concentrated in the 
Earth's core, while Hf became concentrated in the mantle and crust.
If the accretion and separation took place on short time scales (relative
to $\tau_{\rm Hf}$), Hf would still have been radioactive after the 
core-mantle separation, and the final $^{182}$W$/^{183}$W ratio in the mantle
would be higher than the corresponding $^{182}$W$/^{183}$W ratio in 
the original material (often represented by carbonaceous chondritic material,
for example the Allende meteorite), and considerably higher than in the
Earth's core and in iron meteorites. Several people have estimated
the Earth's accretion history based on the $^{182}$Hf$/^{182}$W system. 
Jacobsen (2005)
concluded that the Earth assembled in $\sim 10$\,Myr, and by comparing with
the slightly higher $^{182}$W$/^{183}$W in the lunar material than in the 
present-day reachable part of the Earth's mantle, he found that the 
Earth's core and mantle must have separated (and the Moon formed due to an
impactor of mass ratio relative the Earth of $\sim$ 1:9) already
$\sim$\,30 Myr after the Earth's formation. In agreement with these formation
ideas, un-processed chondritic material, such as Allende, show a 
$^{182}$W$/^{183}$W ratio slightly lower than standard 
terrestrial samples, and iron meteorites even lower. In principle, 
material can therefore be identified as extra-terrestrial if it has
a lower $^{182}$W$/^{183}$W than the Earth's crust. In a study of 
$^{182}$W$/^{183}$W in samples from Isua, 
Schoenberg et al.\ (2002)
found $^{182}$W$/^{183}$W
values two to four sigmas below the standard terrestrial samples, and 
interpreted it as evidence of impact contamination of the Isua crust.
However, the CI (and also the iron meteoritic) 
$^{182}$W$/^{183}$W value is only 
a few times 10$^{-4}$ smaller than the crustal Earth;
$\epsilon_{\rm W}$ = 
[($^{182}$W/$^{183}$W)$^{\rm met}-$($^{182}$W/$^{183}$W)$^{\oplus}$] /
                                  ($^{182}$W$/^{183}$W)$^{\oplus}$
is $-$2.1\,10$^{-4}$ for the Allende meteorite, and
is $-$3.7\,10$^{-4}$ for iron meteorites (and zero for the Earth's 
present-day crust, ($^{182}$W/$^{183}$W)$^{\oplus}$).
This means that the contamination
with cosmic material has to be very large before it will be measurable 
in terrestrial samples. The $^{182}$W$/^{183}$W values reported by
Schoenberg et al.\ (2002)
correspond to a mixture of approximately 50\% of the measured
tungsten coming from Allende-like meteoritic material and 50\% from
terrestrial standard crust (and 1:3 if the cosmic material was iron
meteorites). 

While the negative $\epsilon_{\rm W}$ values do point 
at identified extraterrestrial contamination of the measured material, 
it is not straightforward to use as a quantitative 
measure of the LHB contamination, for the following reasons: 
Bulk cosmic 
material (asteroids or comets) is always associated with a strong 
enrichment of iridium (because iridium is so strongly siderophile that
from a cosmic perspective the Earth's crust has basically zero iridium
abundance). If the tungsten mixing rate of the extraterrestrial and the 
Isua material is representative for the CI to Isua material 
mixing ratio, then the mixed material would end up with an iridium
abundance approximately half that of CI, or well above 200,000 ppt,
which is more than 1000 times above measured values at Isua
(see below). Since the 
measured iridium abundance relates directly to the total 
amount of LHB impacts, such high Ir abundance would also mean that 
the LHB bombardment on Earth should have been $\sim$1000 times stronger
than on the Moon, which has no bearings in dynamics. The meteoritic 
tungsten would therefore have to mix extremely selectively with the 
proto crust in order to in general reach the high mixing values reported.
While this is not impossible, it would leave us with little possibility
to quantify the impact rate or verify that the measured $\epsilon_{\rm W}$
actually represented a general LHB inflow.

Also the chromium isotopic pattern is valuable in identifying 
extraterrestrial matter.
Different types of meteorites have distinctly different chromium isotopic 
patterns, notably the $^{53}$Cr/$^{52}$Cr ratio which is caused by
the decay of $^{53}$Mn which has a half-life of only 3 Myr.
This fact was used by Shukolyukov \&\ Lugmair (1998)
to show that the K/T impactor was likely to be of CI-type
material (i.e.\ a carbonaceous chondrite or a comet). 
However, just as for the $^{182}$W$/^{183}$W ratio, also the 
$^{53}$Cr/$^{52}$Cr ratio in meteorites is only a few sigma different
from the terrestrial standard. In order to identify extraterrestrial
contamination in terrestrial material, it is therefore necessary that 
the amount of extraterrestrial and terrestrial material in the mixture
are of the same order of magnitude (or at least that the amount of 
extraterrestrial chromium in the sample is a large fraction of the 
total amount of chromium). This is certainly the case in 
the K/T dust layer, where $\epsilon_{\rm Cr}$ is almost identical to the 
carbonaceous chondritic value 
($\epsilon_{\rm Cr}$ $\approx$ $-$0.4\,10$^{-4}$), 
and where the chromium abundance itself is close to the chondritic value
(in particular in the K/T dust in Caravaca in Spain where the Cr abundance
is only a factor 2 to 3 below the value in Orgueil and Allende carbonaceous
chondrites). Frei \&\ Rosing (2005)
searched for 
extraterrestrial $^{53}$Cr/$^{52}$Cr anomalies in Isua sediments by use of 
high precision chromium abundance analysis, but found no deviation from
the terrestrial standard. This result is only contradicting the
corresponding $^{182}$W$/^{183}$W results reported in 
Schoenberg et al.\ (2002)
in the 
sense that one study found meteoritic contribution to some of the 
samples studied, and the other didn't. With respect to analysing the 
possible LHB contribution to the Isua proto-crust, both methods 
are not very sensitive to the contamination, and difficult to quantify in 
terms of how much cosmic material might have impacted the area.

Schoenberg et al.\ (2002)
suggested that some kind of weathering processes 
could have been responsible for the up-concentration of tungsten. If a 
similar process didn't affect the chromium abundances, the two studies could 
in principle be in agreement with one another. We, however, conclude that 
the uncertainties and the low sensitivity in the terrestrial material
to cosmic contamination associated with both methods, make them
infeasible to quantify the possible LHB contribution to the 
Earth's Hadean-Eoarchean crust.

In contrast to the tungsten and chromium systems, the abundance of
iridium in the terrestrial crust and in meteorites are widely different
from one another. CI-type meteoritic material, for example, is approximately
20,000 times richer in iridium than the Earth's upper crust,
and even very small contaminations of cosmic material to terrestrial rocks,
will therefore be identifiable.
Hence, we focused specifically on analysis of iridium in this first study
of our samples, and in the analysis of the results we focused in 
particular on finding a unified explanation of the iridium abundances
in the Isua and lunar surfaces (because the same cosmic LHB has 
affected the Moon and the Earth, but given rise to widely different
iridium abundances), and on relating this to the amount of 
impacting material that can be inferred from the lunar crater counting
(because the scaled lunar impact rate must have been accompanied 
by the corresponding amount of iridium that the 
impacting bodies contained).\\

\noindent
{\large \bf Our Isua rock samples.}

The Isua greenstone belt comprises the oldest known major pieces of Earth's 
supracrustal rocks, with an age of $\approx$ 3.8 Gyr 
(Nutman et al.\ 1996, Jenner et al.\ 2008).
The greenstone belt consists of extensive basaltic pillow lava 
flows with intercalated beds of iron-formation, felsic volcanogenic
rocks, pelitic mica schists
and a conglomerate. Intruded into these rocks are mafic and ultramafic
sills and dykes. The whole belt has been repeatedly metamorphosed
under amphibolite facies conditions and suffered several phases
of deformation. The belt has furthermore been intruded by extensive
tonalite sheets. In spite of intense deformation, low strain
domains are frequently seen (Appel et al.\ 1998). In these domains
primary sedimentary and volcanic structures are often seen, such as
conglomerate, well preserved pillows with occelli, pillow breccias
and debris flows. The samples for this study were collected from
banded iron formation, mafic mica schists and from turbidite.

Cosmic impacts completely re-shaped the lunar surface during
the short LHB period, but the contemporary
Isua region has revealed no impact structures at all.
The fact that the Isua sediments represent the oldest known sedimentary
crust, may indicate that it and its surrounding proto-crust can
have been a special place on Earth, relatively quiet and remote from 
major impact craters during the end of the LHB, but Isua may also be a 
completely representative and ''normal" piece of crust, that just happened to 
be the only place not completely re-molten by geodynamical processes during
the subsequent evolution of the Earth until the present day. 
The fact that no craters and shock material have been identified 
may lend support to 
the idea that the Isua proto-crust was particularly far from major LHB
impact craters, at least in the final LHB period.
Whatever the reason is for the lack of impact structures and shocked 
material at Isua, such structures and material would not be the focus 
for understanding the LHB impact on Earth, because they would contain 
information about specific events, and not the sought statistical
information about the bulk LHB impacting on the Earth in general. Instead,
rock types that represent erosion from larger areas of surrounding
land surface, and subsequent sedimentation, will contain the 
bulk mixture of original crust and atmospheric fall-out of
LHB impacted material from the time prior to and during the sedimentation,
which is precisely what one would be most interested in selecting
when trying to understand the influence of the LHB on the early Earth.
By sampling sediments potentially "contaminated" by
atmospheric fall-out and erosional material of the 
land-masses in the hinterlands of the Isua basin (which
likely acted as sediment feeders through weathering and subsequent
transport of dissolved and partiuclate matter by rivers (river loads)),
we therefore do not sample a 
local phenomenon, but are likely to catch the global effect of the last 
part of the LHB on Earth. The sediments of
the Isua greenstone belt are therefore the most likely 
reservoirs on the Earth to potentially reveal representative 
traces of elements that were once part of the LHB impactors.

Chemical sediments that precipitated from the water
(i.e.\ part of the BIFs), on the other hand, 
are not affected by dentrital material transported as river loads.
If the late heavy bombardment 
had still not quite finished while the Isua sediments were deposited,
large impacts will still have brought a mixture of crust and 
impactor material into the stratosphere
in the form of dust that later will have rained out of the atmosphere. 
Dust from this fall-out will mechanically have mixed with 
any sediments in open waters, including material from chemically
precipitated sediments.
If sediments from both river load erosion and chemical precipitation include 
traces of cosmic material, it will therefore be an indication that 
the LHB was still in its final phases on the Earth 3.8 Ga ago, in 
agreement with what we should expect from scaling of the Apollo 
dating of lunar impact melts to the conditions on Earth.

We have therefore sampled three different types of metasedimentary rocks:
(1) shallow water pelagic sediments (garnet-bearing mica-schists) with 
detrital input from both mafic volcanic and intrusive rocks
(Bolhar et al.\ 2004),
(2) clastic sediments (turbidites) deposited in deep water
(Rosing 1999),
and (3) detritus-free oxide-facies banded iron stones (BIF) characterised by 
alternating magnetite-rich and silica-rich microbands
(Polat \&\ Frei 2005; Frei \&\ Polat 2007)
Age determinations of the region constrain the deposition of the sediments 
to $\sim$\,3.8 Ga ago (Nutman et al.\ 1996).
The sediments are composed of eroded material derived
from the contemporaneous surrounding land surface
as well as possible atmospheric fall-out.

Trace element systematics of Isua greenstone belt metasediments
show strong resemblance to other well-documented Archean clastic
sediments, and are consistent with a provenance consisting of
ultramafic, mafic and felsic igneous rocks (Bolhar et al.\ 2005).
Major element systematics document incipient-to-moderate source
weathering in the majority of metasediments, while signs of secondary
K-addition are rare. Detailed inspection of Eu/Eu$^*$, Fe$_2$O$_3$
and CIW (chemical index of weathering) relationships revealed that
elevated iron contents (when compared to average continental crust)
and strong relative enrichment in Eu may be due to precipitation of
marine Fe-oxyhydroxides during deposition of diagenesis on the 
sea-floor (Bolhar et al.\ 2005).

The voluminous mafic volcanic rocks are composed primarily of
pillow basalts intercalated with ultramafic units. Banded iron
formation, cherts, conglomerates and siliciclastic turbidites
are intercalated. The geochemical characteristics and features
represent a coherent mafic to ultramafic suite, comparable to those
of Phanerozoic boninites (Polat et al.\ 2002).
Given the observation that in the Tertiary, boninites are 
exclusively associated with intra-oceanic subduction environments
(e.g.\ Izu-Bonin-Mariana subduction system), this suggest that
the Isua metabasalts were formed in intra-oceanic subduction
zone-like geodynamic processes (Polat et al.\ 2002), 
although there is still an ongoing debate as to the existence
of subduction zones this early in the Earth's history (Glikson 2004).

Geochemical, lithological, and structural data from the
Isua greenstone belt are all collectively consistent with a 
convergent margin geodynamic setting. The Isua BIFs are
spatially and temporally associated with boninitic and island
arc picritic pillow basalts.
Given the observations that Cenozoic boninites and picrites tend
to form in an intra-oceanic arc-forearc setting, Polat \&\ Frei (2005)
suggested that the Isua BIF-boninite and BIF-picrite associations
were also deposited in a similar geodynamic setting. 
The high-temperature hydrothermal alteration of the oceanic crust 
produced significant  hydrothermal discharge with large 
quantities of Fe necessary to contribute to the deposition
of the Isua BIFs. The hydrothermal alteration may have resulted
from the opening of an asthenospheric window developed as a 
consequence of ridge subduction beneath an early Archean
arc-forearc region (Polat \&\ Frei, 2005).
The ridge subduction model can also explain the origin of the 
contemporaneous tonalite-trondhjemite-granodiorite (TTG) 
intrusions in the Isua region.
Partial melting of laterally accreted and thickened oceanic crust 
under amphibolite to eclogite metamorphic conditions by 
upwelling of asthenospheric windows may have produced TTG melts.\\

\noindent
{\large \bf The iridium abundance analyses.}

\begin{table}
{\footnotesize
\begin{tabular}{|lr|lr|}
\hline
\hline
  Identification   &  Concentration & Identification   &  Concentration \\
                   &  ppt           &                  &  ppt           \\
\hline
 mica-schist       &                & BIF, silica-rich&            \\
                   &                & microbands      &            \\
 491061-1          &  247$\pm$29    & 3062-Q1         & 13$\pm$7   \\
 491061-2          &  140$\pm$24    & 3062-Q2         & $<$5       \\
 491061-3          &  120$\pm$32    & PA1-1-Q         & 35$\pm$8   \\
                   &                & PA1-2-Q         & 22$\pm$6   \\
                   &                & PA2-1-Q         & 492$\pm$40 \\
                   &                & PA3-Q1          & 44$\pm$4   \\
                   &                & PA3-Q2          & 9$\pm$2    \\
                   &                & PA4-1Q          & 140$\pm$7  \\
                   &                & PA4-2Q          & 18$\pm$3   \\
                   &                & 491232B(2)      & 433$\pm$20 \\
                   &                & 491232D         & 85$\pm$10  \\
 simple average    &  169$\pm$16    &simple average   & 106$\pm$5  \\
\hline
 turbidites        &                & BIF, magnetite- &           \\
                   &                & rich microbands &            \\
 810216(2)         &  101$\pm$14    & 3062-M1         & $<$50        \\
 810217            &  $<$6          & PA1-1-M         & 170$\pm$30    \\
 810194            &  21$\pm$8      & PA1-2-M         & $<$40    \\
 460528            &  $<$19         & PA1-3-M         & $<$100   \\
 810208            &  $<$10         & PA2-1-M         & $<$100   \\
 Bouma-1           &  250$\pm$40    & PA3-M1          & $<$70    \\
 810194(2)         &  140$\pm$20    & PA3-M2          & $<$70    \\
 810207(2)         &  100$\pm$10    & PA4-1M          & 100$\pm$20    \\
 810208(2)         &  6850$\pm$160  & PA4-2M          & $<$80    \\
 460534(2)         &  231$\pm$12    & PA4-3M          & $<$90    \\
 460528            &   86$\pm$12    & 491232A         & $<$70    \\
 simple average    &   96$\pm$7     & 491232C         & $<$60    \\
 weighted average  &  185$\pm$8     & simple average  & $<$80    \\
\hline
\end{tabular}
\caption{Measured iridium concentration in ppt in three types
of sediments from the Isua greenstone belt, Western Greenland.
A '(2)' after the sample-name means that two separate sub-samples were
measured individually, and the average is reported here. 
}
\label{table1}
}     
\end{table}
In Table 1 we report our results of high-precision iridium
abundance measurements in 37 individual samples, grouped 
according to rock type. 
Mica-schists and turbidite samples were crushed in an agate mortar
from whole rocks chips. BIF samples from quartz- and magnetite-rich
mesobands were separately crushed and analysed.
When possible, we divided our samples into two or more sub-samples, 
which were then measured separately in order to trace possible larger
inhomogeneities in the sample.  The iridium
abundance in the sub-samples were typically within a factor of two from
one another.  One of the turbidity current samples, however, seemingly 
had a nugget resulting in an extremely high iridium concentration of 
13700 ppt in one of the sub-samples, and the reported 
value is an average (6850 ppt) of this and a successive measurement
of the other sub-sample of this same sample. The listed average of the
turbidites is calculated without including the nugget value.
Regrettably, we obviously know nothing 
more about this assumed nugget, since nothing particular was noticed 
during the mortaring of this particular sample.
Also listed is the weighed average (by weight) of the individual samples.

Mortared samples were irradiated at the {\sc jaea}'s research reactor, JRR-3,
without prior chemical separation.
Gamma-gamma coincidence spectroscopy
(Oshima et al.\ 2002, Hatsukawa et al.\ 2002, Toh et al.\ 2001,
Oshima et al.\ 2008),
of multiple $\gamma$-rays from
the radio-isotopes produced by the neutron capture reactions, were then 
performed with an array of twelve Ge detectors equipped with BGO Compton
suppressors GEMINI-{\sc ii}
(Oshima et al.\ 2008).
Each sample was typically of 50$-$100 mg, and were measured 
for about 24 hours after 4 weeks of radiation. This method has previously 
been demonstrated
to be capable of measuring iridium concentrations as low as a few ppt with
a relative accuracy of a few percents 
(Oshima et al.\ 2002, Hatsukawa et al.\ 2002, Toh et al.\ 2001,
Oshima et al.\ 2008, Hatsukawa et al.\ 2003, Hatsukawa et al.\ 2007).

It is seen that the mica-schists and turbidites consistently
have a bit higher Ir enrichment than the BIFs, and 
we take their value as representative for the surrounding solid crust prior to
the erosion and following deposition of the sediments. Magnetite-rich
BIF samples are more difficult to measure than the silica-rich BIF, 
because they give rise to 
a higher Ir-detection limit. We envision that the BIF has formed as
chemical precipitate in the ocean with addition of a cosmic fall out
(analogue to the K/T-fall out) during the period of sedimentation. 
We therefore conclude from Table 1 that the pre-existing Hadean-Eoarchean
crust just prior to the depositional period that led to the sedimentations
we now see at Isua, had an
iridium abundance close to the value of the mica-schists and turbidites,
and that there still was an atmospheric cosmic fall out in the ocean
at the time of BIF sedimentation -- all in qualitative agreement with our 
expectations from a declining LHB and its lunar records prior to the 
formation of the Isua sediments.
We therefore conclude from Table 1 that a rough round number for 
the cosmic iridium contribution seen in the sediments is 150$\pm$25 ppt.\\

\noindent
{\large \bf The source of the iridium in the sediments.}

In order to test to which extent the iridium concentrations in the
metasediments of the IGB are controlled by detrital (i.e.,
particle-controlled) components, we need to distinguish between chemical
sediments (BIF microbands) and clastic sediments (mica-schists, turbidites).
Based on
chemical (major and trace element) and isotope geochemical (Pb isotope) 
characterisation of ancient clastic and chemical sediments from the IGB
(and TTG intrusions emplaced within the IGB)
Kamber et al.\ (2005) and Bolhar et al.\ (2005) concluded that the Isua
protocrust probably had a mafic character and showed a slightly enriched
signature which these authors attributed to an earlier (i.e., pre-3.8
Ga) differentiation-recycling event. In the view of this scenario, and
based on the exposure itself of mafic volcanic rocks (i.e., boninitic
and picritic basalts; Polat et al., 2002; Polat et al.,2003) within the
IGB, we discuss the measured Ir concentration in the samples presented
herein as follows:\\

\noindent
{\bf Chemical sediments (BIFs)}

BIFs from within the IGB are characterised by low detrital components
as shown by Frei \&\ Polat (2007). These authors used scandium
concentrations to monitor the lithogenic element concentrations in IGB
BIFs. Similar to other such studies which used Sc concentrations of $<$20
ppm to reflect very low detrital components in BIFs (e.g., Alexander et
al., 2008), Frei \&\ Polat (2007) argued that Sc concentrations smaller
that 1 ppm in the Isua BIFs indicate very pure chemically precipitated
sediments, which they support by REE patterns which resemble present-day
seawater. These values are in agreement with very low Al$_2$O$_3$ contents of
$<$0.5 wt\% (our unpublished data). On the basis of these investigations,
the Ir concentrations in silica-rich BIF mesobands given in Table 1,
are interpreted to reflect an extraterrestrial contribution to
the sediment. It is difficult to assess in which form this Ir may have
co-sedimented, i.e., whether Ir is hosted by minor particles or
precipitated from an initially dissolved form. The very low Al$_2$O$_3$ and
MgO concentrations (both $<$0.5 wt\%; our own unpublished data) in these
BIFs argue against a small mafic particulate component (either eroded
from the continental hinterland or direct fall-out from the atmosphere)
that could have co-sedimented with the silica-rich bands. Even if we 
assumed that as much as 0.5wt\% MgO derived from basaltic precursors
(such as boninites and picrite typical of the IGB) with Ir
concentrations comparable to Archean komatiites, mafic komatiites and
basalts (MgO from 5-20 wt\%) with up to 1 ppb Ir (Hong et al., 2006;
Puchtel and Humayun, 2000; Maier et al., 2003; and others), then the Ir
concentrations in such mesobands should not exceed $\sim$25 ppt. We therefore
interpret the elevated Ir concentrations in these BIF mesobands to
derive from extraterrestrial sources (mainly as atmospheric
fall-out) with high Ir concentrations.
The iron-rich mesobands of the BIFs studied herein, despite the
elevated analytical detection limits for Ir, are characterised by Ir
concentrations that are similar or lower than the respective
concentrations in the silica-rich mesobands (cf.\ Table 1). Since these
mesobands are demonstrably dominated by hydrothermal (vent-derived)
input (high Eu anomalies; Frei \&\ Polat, 2007) with presumably very low
dissolved PGE, the average limit of detection around 80 ppt is regarded
as a maximum for the Ir levels in these horizons, which compares well
with the $\sim$100 ppt average of the silica-rich mesobands, and
consequently interpreted by us to indicate admixture of the same
extraterrestrial component into the chemical sediment.\\

\noindent
{\bf Clastic metasediments}

Similar calculations for the interpretation of Ir levels measured in
the clastic metasediments from the IGB may indicate a smaller
mafic proto-crustal addition to
the source of PGE in these lithologies. Bolhar et al.\
(2005) showed that trace element systematics of IGB metasediments
strongly resemble other well-documented Archean clastic sediments, and
trace elemental signatures are consistent with a provenance from mixed
ultramafic, mafic and felsic igneous rocks. In particular, based on
Ti/Zr vs.\ Ni concentration diagrams, these authors propose that IGB
metasediments are broadly consistent with approximately equal
proportions of average basalt and boninite admixed to average felsic
volcanogenic sediment and TTG. 
Minor additional mixing of a komatiite-derived clastic components
with Ir concentrations as high as 1.0$-$1.5 ppb Ir (Puchtel \&\ Humayun,
2000; Anbar et al., 2001; Maier et al., 2003) cannot be excluded without
future additional extensive studies of isotopic (in particular osmium) and 
elemental abundances. However, the most important observation
at this place from the results presented in Table 1, is:
(1) the remarkable similarity in the iridium abundance in the three completely
different types of sediments, (2) the rough average of a factor seven
iridium excess compared to present day ocean crust and upper continental
crust, and
(3) that none of the sediment types have an iridium excess much higher
than the rough average of 150\,ppt. The last point is very important, as 
will become clear from the following discussion, because an 
asteroidal dominated LHB would have resulted in considerably higher
iridium excesses in the Isua sediments, so even if smaller amounts of
iridium-rich komatiite-like clastic components would have mixed into
the measured clastic metasediments, this could admittedly add a complication 
in the understanding and quantifying of the derived average iridium excess, 
but it would 
not change the conclusion about a cometary dominated LHB impact.\\

\noindent
{\large \bf Iridium and the Earth-Moon formation.}

A standard theory for the formation of the Earth-Moon system
(Martin et al.\ 2006),
is that most of the Earth accreted rapidly from 4.57 Ga to 4.5 Ga ago.
Near the end of this period the core and the mantle had separated,
when a Mars-sized object collided with the proto-Earth, and resulted in
the Moon forming from expelled mantle material (Jacobsen 2005,
Martin et al.\ 2006). At this time the
upper Earth and the Moon obviously had an identical abundance of
iridium (and all other refractive elements).
The Moon never totally differentiated, but the upper $\sim$100 km 
melted and homogenised to form the present feroan anorthosite surface. 
The oldest Moon dust is 4.42 Ga (Snyder et al.\ 2000, Ryder 2002), 
and the outermost lunar crust must 
therefore have been solid since that time. Almost all the craters we 
see on the Moon today were, however, formed during the narrow LHB period from 
$\approx$\,3.9\,Ga to 3.8\,Ga ago. The iridium abundance in the present lunar
surface is therefore coming from material of composition close to the
Earth's early mantle (i.e., very low in Ir)
mixed with cosmic impacts from the LHB period (plus 
possible impacts from the period from 4.42 to 3.9\,Ga ago). Likewise
the Earth's proto-Isua crust is a mixture of this same first Earth-mantle
material and material from the same kind of impactors that formed 
the lunar craters. The lunar anorthosite
has $<$\,10\,ppt Ir (Lodders \&\ Fegley 1998, Wedepohl 1969)
(while unprocessed meteoritic material such as
Allende have 465,000\,ppt, and the Earth's present-day upper crust and 
ocean crust have 20\,ppt). The fact that we find that the Isua proto-crust 
had an Ir abundance higher (15 times or more) than the lunar anorthosite,
shows that the cosmic impactors that hit both the Earth and the Moon during
the LHB, were of such a character that they deposited their Ir on the 
Earth but essentially not on the Moon. 
This might at first seem intuitively impossible, 
but in fact it instead contains crucial information about the impactors and,
in particular, their velocity.
Our numerical models, described below, 
show that the measured difference in Ir deposition
would occur if the impactors were comets, but not if they were 
asteroids (``meteorites").  The basic theory and assumptions 
are similar to what has been adopted by others
(e.g.\ Chyba 1991, and Melosh \&\ Vickery 1989),
but here applied to estimate the expected abundance of iridium.\\

\noindent
{\large \bf Asteroid versus comet crater formation and crustal enrichment.}

The total mass that impacted the Moon during the period in which today's
visible craters were formed, can be 
calculated by transforming the observed crater sizes to impact mass.
The transformation depends on several parameters, noticeably the
energy per impacting mass unit, which is proportional to
the impacting velocity squared, $v_i^2 = v_\infty^2 + v_{esc}^2$, where 
$v_\infty$ is the impactor velocity far away from the Earth-Moon, and $v_{esc}$
is the escape velocity (from Earth or Moon). 
Since $v_\infty$ is higher for comets than for asteroid impactors,
the total impact mass necessary to form the lunar craters, 
is somewhat smaller if they were comets than if they were
asteroids, but typical values reached 
in the literature are around 10$^{20}$ kg
(Martin et al.\ 2006, Chyba 1991, Hartmann et al.\ 2000).

The impacting mass on the Moon can be scaled to corresponding impacting mass
on the Earth. 
The effective gravitational area, $A_{\rm eff}$, that an approaching
impactor will see, is related to the geometrical area, $A_{\rm geom} 
= \pi \, r^2$, by
\begin{equation}
A_{\rm eff} = A_{\rm geom} (1 + 2\theta)
\label{gravfac}
\end{equation}
where the proportionality factor, or gravitational enhancement factor, 
$(1+2\theta)$, is given by the Safronov number
\begin{equation}
\theta = \frac{G m}{r v_\infty^2} \ = \ E_{pot}/2E_{kin}
                   \ = \ {1\over 2}\left({v_{esc}\over v_\infty}\right)^2
\label{safronov}
\end{equation}
While the escape velocity, $v_{esc}$, the target mass, $m$, and the
geometrical radius, $r$, (of Earth and Moon) are known, 
it is not obvious what $v_\infty$ was,
and it depends on the assumption of the nature of the impactors. 
Comets will, as a group, have more elliptical orbits, and cross the Earth's 
orbit with higher velocities than asteroids.
By assuming that typical asteroids
had $v_\infty \approx$ 12 km/s (as present-day Earth-crossing asteroids)
and typical comets had  $v_\infty \approx$ 20 km/s (as present-day 
short-period comets), we can express the ratio of impacting mass per
m$^2$ on the Earth and the Moon, $m^\oplus /m^{\rm moon}$, as
\begin{eqnarray}
\label{imprat}
m^\oplus / m^{moon} &=& 
          \left( {A_{\rm eff}^\oplus \over A_{\rm eff}^{moon} } \right)
    \left( {r^{moon} \over r^\oplus} \right)^2  \nonumber  \\
        &=& (1 + 2\theta^\oplus) / (1 + 2\theta^{moon}) \nonumber \\
        &=& \left( {v_i^\oplus \over v_i^{moon}} \right) ^2 \\
    &=& 1.8\ {\rm for\ asteroids\ and\ }  1.3\ {\rm for\ comets} \nonumber
\end{eqnarray}
We see that with the appropriate data for Earth and Moon used in 
Eq.\ \ref{safronov} and \ref{imprat}, 
and the velocity estimates given above, 
we can scale the lunar cratering rate to Earth impacts, and conclude
that 1.8 times more mass must have hit 
the Earth than the Moon per m$^2$ surface if the impactors were asteroids,
and 1.3 times more if the impactors were comets.

The corresponding ratio, $f^\oplus / f^{moon}$, of 
impacting energy per m$^2$ of the Earth ($f^\oplus$) and 
the Moon ($f^{moon}$) is,
\begin{eqnarray}
\label{fluxrat}
f^\oplus / f^{moon} &=& {m^\oplus \over m^{moon} } \,
    \left( {v_i^\oplus \over v_i^{moon}} \right)^2  \nonumber  \\
        &=& \left( {v_i^\oplus \over v_i^{moon}} \right) ^4 \\
    &=& 3.2\ {\rm for\ asteroids\ and\ }  1.7\ {\rm for\ comets} \nonumber
\end{eqnarray}

As a side remark we see that the ratio of total impacting mass 
hitting the Earth and Moon is
\begin{equation}
\label{totmassrat}
\left( {m^\oplus \over m^{moon}} \right) 
               \left( {r^\oplus \over r^{moon}} \right)^2 
    \approx 24\ {\rm for\ asteroids\ and\ }  18\ {\rm for\ comets}
\end{equation}
(since $( r^\oplus / r^{moon} )^2$ = 13.5).

In reality, $( m^\oplus / m^{moon} )$ may well have been $\approx$
a factor two larger than given by Eq.\,\ref{imprat}, 
because the calculations leading to Eq.\,\ref{imprat} only includes
the effect of impactors of the same size hitting the Earth and Moon,
while considerations about small number statistics of the 
very largest impactors, show that the larger probability of a body
hitting the Earth, implies that the few very largest impactors (larger
than the one that formed Mare Imbrium on the Moon) will 
most likely have hit the 
Earth and added considerably to the ratios in Eq.\,\ref{imprat}.

The enrichment the impacting material will create in the crust depends 
not only on the total impacting mass and its composition, but also
on the ratio between mixed and re-emitted material during the impact.
It happens 
that this ratio is very different for Earth and Moon, and different 
if the impactors were comets or asteroids.
This is basically because the LHB impactor velocities were comparable to 
the escape velocity of Earth, but much larger than the escape 
velocity from the Moon. This is a situation very different from
that of the accreting proto-planets, where the relative velocities
of the accreting and accreted material were small.
The quantification of the different ratios between accreted and
re-emitted material contain important clues to understanding what
type of impactors were causing the LHB.
The ratio of escaping to mixing material 
scales with the number of impactors that form plumes of high 
speed material (i.e.\ with $v>v_{\rm esc}$), 
relative to those that don't, which
is a function of impactor composition, impactor velocity distribution
($v_\infty$), and target escape velocity ($v_{esc}$). 

In order to create a vapour plume that expands with more than the escape 
velocity, obviously the impacting energy must exceed the energy required 
to evaporate the impactor (and other material to be included in the plume)
plus accelerate this mass of material to above the escape velocity.
By assuming that such a plume-creating impact will evaporate and 
carry away the impactor itself plus an identical amount of target
mass, and that the downward absorbed energy is identical to the 
upward released energy in the gas expansion (Melosh 1989,
Melosh \&\ Vickey 1989), we can express
the minimum impact velocity, $v_{min}$, required from
\begin{eqnarray}
\label{vminplume}
{1\over 2} ( {1\over 2} m v_{min}^2 ) &=& 
                  2 ({1\over 2} m v_{esc}^2) + 2 m H_{vap}
                                      \ \ \Leftrightarrow   \nonumber \\
v_{min}^2 &=& 4 (v_{esc}^2 + 2 H_{vap})
\end{eqnarray}
where $H_{vap}$ is the vaporisation energy (the enthalpy of 
vaporisation), which,
expressed in MJ/kg (=km$^2$/s$^2$), is
\begin{equation}
H_{vap} = 13 {\rm MJ/kg\ for\ silicates} \ \ \ \
H_{vap} =  3 {\rm MJ/kg\ for\ ice}
\end{equation}
which with $v_{esc}^{moon}$=2.4 km/s, and $v_{esc}^\oplus$=11.2 km/s gives
\begin{eqnarray}
\label{vminval}
v_{min}^{moon} &=& 11\ {\rm km/s\ for\ asteroids,\ and\ 7\ km/s\ for\ comets} \nonumber \\
v_{min}^\oplus &=& 25\ {\rm km/s\ for\ asteroids,\ and\ 23\ km/s\ for\ comets}
\end{eqnarray}
With the velocity distributions from Earth crossing asteroids and 
Earth-crossing short-period comets, one concludes from
Eq.\,\ref{vminval} 
that on the Moon approximately 50\% of asteroids and 100\% of comets will 
create Moon-escaping plumes, 
while the corresponding numbers for the Earth are 10\% and 50\%.
On the Moon the plumes will result in that all the plume mass
is lost, while on Earth this is only the case if the plume
mass (with velocity above $v_{esc}$) is larger than the mass of the "hat" 
of atmosphere above and tangentially to the side of the impact (which
is then also lost into space).

Apart from the mass that is lost with the plume, also direct ejecta from 
the impact crater can be lost. Chyba (1991) found that the mass of the crater 
material, $M_{ejc}(v>v_{esc})$, 
ejected with velocity $v$ above $v_{esc}$ expressed as function 
of density $\rho$ and mass $m$ of the impactor, and density $\rho_t$ of the 
target, can be expressed as
\begin{equation}
\label{mejcti2}
M_{\rm ejc}(v>v_{esc}) = 0.11 \left({\rho \over \rho_t}\right)^{0.2}
                     \left({v\over v_{esc}}\right)^{1.2}m
\end{equation}

Introducing for the Moon ($\rho_t$\,=\,2.9\,g/cm$^3$) $v_{med}\,=\,$12\,km/s
and $\rho$\,=2.2\,g/cm$^3$ for collisions with asteroids and 
$v_{med}$\,=\,20\,km/s and 
$\rho$\,=\,1.0\,g/cm$^3$ for comets, 
we find the mass of material ejected with velocity larger
than $v_{esc}$ (=2.4km/s) to be  $M_{ejc}(v>v_{esc})$ = 0.7$m$ for 
asteroids and 1.1$m$ for comets. 
I.e., for asteroids 70\% the impact mass is escaping the Moon 
in the form of directly ejected lunar surface rocks, and for typical
comets the same amount (i.e., $\approx$ 1.1\,$m$) 
of crater material escapes the Moon as the mass of the impacting comet.

An "average" asteroid of mass $m$ colliding with the Moon, will 
therefore (according to the above formulas and numbers introduced)
cause the Moon to accrete half of the asteroid mass, 
lose  the other half in a plume, lose  additional
$0.5m$ worth of lunar material in the plume, 
plus loosing 0.7$m$ of lunar material in crater ejecta material. 
Net, we therefore see that the Moon will become 0.7$m$ lighter for each
heavy bombardment asteroid impact of mass $m$. 

An "average" comet of mass $m$ colliding with the Moon, will 
cause the Moon to lose  all the comet mass in a plume, lose  additional
$1\,m$ worth of lunar material in the plume, 
plus loosing 1.1$m$ of lunar material in crater ejecta material. 
Net, we therefore see that the Moon will become 2.1$m$ lighter for each
heavy bombardment cometary impact of mass $m$. 

While we see from the above that a typical comet will leave no traces 
apart from the crater (because all the cometary material escapes the Moon
in the plume), the crust will change composition if the impacts were
asteroids, because net the Moon will accrete 0.5$m$ of (chondritic) meteoritic
material (100\% enriched in chondritic material), while it will lose  
1.2$m$ lunar material enriched with (today) 2\% chondritic material, 
so net gaining at least 0.5$m$-0.02$\times$1.2$m$=0.48$m$$\approx$0.5$m$ 
chondritic material per $m$ chondritic impact.
The net loss of material is therefore in agreement with an increase of 
chondritic material in the crust during the late heavy bombardment
with asteroids.

The corresponding numbers for the Earth give that 10\% of asteroids and 50\% of comets
will create a plume, implying that on average 0.2$m$ will leave the Earth in plumes per asteroid
impacts of mass $m$, and 1$m$ crater and comet material will on average be ejected when a 
1$m$ comet impact the Earth.
An "average" speed asteroid (i.e., $v$\,=\,$v_{med}$\,=\,15\,km/s) 
of mass $m$ will accelerate
0.14$m$ crater material to velocities above the Earth's escape velocity, 
while a corresponding comet (i.e., $v$\,=\,$v_{med}$\,=\,23\,km/s) will 
eject 0.20$m$ crater material. On average
therefore $m$ mass asteroids will leave 0.9$m$ chondritic material on Earth and 
eject 0.1$m$ asteroid plus 0.25$m$ Earth mantle material, thereby making
the Earth 0.55\,$m$ heavier. 
A cometary impact of mass $m$ will inject 0.5$m$ cometary material into 
the Earth's crust, return 0.5$m$ comet to space
together with 0.7$m$ terrestrial material, thereby making the
Earth 0.2\,$m$ lighter. 

As opposed to the Moon, the Earth is likely to accrete some of the ejected 
material again later in a new collision, possible at lower encounter velocity,
while the ejected material from the Moon has a higher probability to
be accreted by the Earth than by the Moon after entering an Earth crossing 
heliocentric orbit.

In summary, we find that on the Moon only half of the 
impacting asteroids, but almost all of the comets, will form high-velocity 
plumes. On Earth the corresponding numbers are 10\% for asteroids and 50\% 
for comets (taking into account also that on Earth
the plume energy has to be large enough to blow away the atmosphere above it 
before it can disappear into space). 

For typical comets, the above numbers therefore show that 50\% of their 
impact mass would mix with the 
terrestrial crust (and 50\% being lost in plume ejection), while 
on the Moon all of a comet would be lost into space in plume ejection.
Therefore an impacting comet will leave Ir on Earth but not on 
the Moon. In contrast, we have seen that typical
asteroids hitting the Earth and the Moon will leave 
respectively 90\% and 50\% of their mass mixed with the crust.
Therefore asteroids will leave Ir on Earth as well as on the Moon.

The final iridium abundance of course also depends on how long time 
this impacting took place on solid crust, and how stiff the crust was
relative to the impacting energy, as function of time.
If the Moon and Earth were identical bodies in this respect,
an impacting asteroid (of a certain impact energy) 
would leave 9/5 = 1.8 times more Ir on the Earth 
than on the Moon. Several factors contribute to lowering this 
number (lower volume to surface ratio and smaller impact energy 
contribute to a faster cooling and solidification of the lunar crust, 
and higher 
escape velocity on Earth contributes to deeper mixing of impactors 
in otherwise identical crusts). In spite of this, we do not find a
smaller Ir abundance, or even a comparable Ir abundance, but rather a 
substantially higher (a factor 15 or more) Ir abundance in our Isua 
samples than what is known from the lunar anorthosite.
We therefore conclude that the impactors cannot have been asteroids.

Comets, on the other hand, are in qualitative agreement with finding almost
no Ir in the lunar anorthosite (because of 100\% plume escape), 
and at the same time enhanced Ir in the Isua samples. 
We have therefore shown above that the relative lunar and Isua
iridium abundances are in qualitative agreement with a cometary
LHB, but not with an asteroid LHB.
We will now argue that the amount of Ir we list in Table 1, within
reasonable basic assumptions, not only is in 
qualitative, but also in quantitative,
agreement with the expectations from cometary dominated LHB impacts. \\

\noindent
{\large \bf The LHB iridium deposition on Earth.}

In order to quantify the effect of the LHB impacts, we first need to 
quantify what is meant by the LHB, which has come to have a 
rather diffuse meaning in the literature.
Strictly speaking, the transformation from lunar crater counts
to impact mass only gives us a lower limit for the total mass of the 
LHB impactors. The lunar surface is saturated with craters, and 
the age of the oldest measured impact melts therefore gives us the 
age at which the impacting energy density for the last time dropped 
below the saturation point. Any impact melted rock older than
this age will per definition have been hit by a new impact later, 
thereby being re-melted and having its melting age reset. 
The peak of the LHB is defined as this saturation age. The fact that
the energy density of impacts could have been even larger before the 
LHB peak, is the root of the well known discussion of whether there
at all were a late heavy bombardment (or we just see the 
post-saturation declining part of a much stronger impact flux that 
possibly was the tail of the accretion that formed the planets and
moons). The reasons for believing that there was an LHB include:
(1) that the $\approx$ 10$^{13}$ comets estimated to now be in 
the Oort-cloud must have been perturbed in random directions
from their formation place in the plane between Jupiter and Neptune
some time after the formation of the solar system (and some of these
must necessarily have hit the inner planets and moons in some kind of 
time-limited bombardment at that time), 
(2) that it is theoretically difficult to 
envision that the planetary accretion process could have taken as long as 
700 million years until the end of the LHB, and
(3) that the impacts that created the basins on the Moon may have 
been as abundant as $\sim$1 per 10 Myr around the LHB peak, but then 
abrubtly ended $\sim$3.8 Gyr ago.

The amount of iridium one would expect to have mixed into 
the proto-Isua crust as a result of the LHB, obviously
depends on the assumption that there actually was a LHB, 
and it also depends on how long time such a bombardment took place
with the capacity to mix material into the surface rocks.
While the first assumption (that there was a LHB) is very likely to 
be correct (among other things
for the 3 reasons given above), the latter is unknown and not likely
to be answerable, at least in the foreseeable future.
We will assume that the cosmic material which mixed into the original 
crust material, both on the Moon and on the proto-Isua crust, came
from the LHB peak and onward in time. While this assumption 
affects the quantification of the expected iridium abundance at Isua,
it does not affect the question of whether the impactors were
comets or asteroids.

The Apollo dating of the lunar impact melts has been
fitted in the literature with an exponentially decaying impact rate,
\begin{equation}
\label{n(t)}
N(t) = N(t_0) e^{-(t-t_0)/\tau} = 
       N(t_0) e^{-ln2(t-t_0)/\tau_{1/2}} 
\end{equation}
with a half life, $\tau_{1/2}$, between 30 and 100 Ma 
(Chyba 1991).
In Eq.\,\ref{n(t)}, $t$ and $t_0$ could in principle be any 
two times during the LHB, but we will conveniently think of a 
normalisation such that $t_0$ is 
the time of the LHB peak (and hence $N(t_0)$ the number density of
impacts at the LHB peak) and $t$ any time later than $t_0$.
The LHB peak (in the way it was defined above) occurred somewhat 
later on the Earth than on the Moon (but the decay rate, $\tau$ or
$\tau_{1/2}$, will have been the same). This is because the
energy per mass unit impactor per m$^2$ is higher
on Earth than on the Moon, and the craters are explosion features,
such that their diameters scale roughly with the 3$^{\rm rd}$ root
of the impacting kinetic energy density, $f$ in Eq.\,\ref{fluxrat} 
(not the mass density $m$ in Eq.\,\ref{imprat}), such
that the saturation point
will have occurred on the Earth (i.e. the terrestrial LHB peak)
only when the terrestrial impacting flux, $f^\oplus$(t),
had fallen to the value $f^{moon}(t_0)$
it had at the Moon during the lunar LHB peak.
As the kinetic energy depends on whether the impactors were
asteroids or comets, the time of the terrestrial LHB peak also
depends on the nature of the impactors, and has to be evaluated
independently in the two cases.
If large areas of terrestrial crust from the time of the LHB still
existed, the time of the terrestrial LHB peak could have been defined 
empirically from crater counts and crater dating, just as it is 
done for the Moon. It is because such pieces of the terrestrial crust
do not exist, that we are forced to scale the lunar LHB peak instead,
and introduce the associated additional challenges. 

The impacting mass flux as function of time can be calculated from
Eq.\ref{n(t)} and the kinetic energy flux ratio on the Earth
and Moon at any given time can be computed from Eq.\ref{fluxrat}.
We can therefore find the delay, $\Delta t = (t-t_0) / \tau_{1/2}$, 
in the terrestrial LHB peak compared to the lunar LHB peak, 
by expressing the demand of $f^\oplus(t) = f^{moon}(t_0)$ from
these two equations, such that
\begin{eqnarray}
\label{lhbdelay}
 {N(t) \over N(t_0)} &=& 
                  \left( {v_i^{moon}(t_0)\over v_i^\oplus(t)}\right)^4
             \ = \  \left( {v_i^{moon}(t)\over v_i^\oplus(t)}\right)^4
                                      \ \ \Leftrightarrow   \nonumber \\
    e^{-ln2 (t-t_0)/\tau_{1/2} } &=& 
           \left( {v^2_\infty + (v_{esc}^{moon})^2 \over 
                    v_\infty^2 + (v_{esc}^\oplus)^2} \right)^2
                                      \ \ \Leftrightarrow    \\
    {(t-t_0) \over \tau_{1/2} } &=& 
         {ln(v_i^\oplus / v_i^{moon} )^4 \over ln2}  \nonumber  \\
    &=& 1.7\ {\rm for\ asteroids\ and\ 0.75\ for\ comets} \nonumber
\end{eqnarray}
If the lunar LHB peak ($t_0$ in Eq.\,\ref{lhbdelay}; usually just referred 
to as the LHB peak) was at $t_0$ = 3.9 Ga and the half life was 100 Ma,
then the terrestrial LHB peak was at 3.825 Ga for a cometary LHB and
at 3.73 Ga for an asteroidal LHB. It would then be very 
surprising that we find no impact structures and shocked material 
at Isua (in particular if the impactors were asteroids).
If the lunar LHB peak was at 3.95 Ga and the half life 30 Ma, 
then the terrestrial LHB peak was at 3.93 Ga for a cometary LHB and
at 3.90 Ga for an asteroidal LHB. There would then 
have been almost no more impactors left at the time the
sedimentation occurred that would become Isua, and in that
case at least the chemically precipitated minerals (i.e., part of the BIF) 
would show no trace of enhanced iridium.
The correct values of $t_0$ and $\tau_{1/2}$ therefore must be somewhere
in-between, in order to agree with both the morphological and 
the geochemical data (and the corresponding lunar data).

The fraction $F_{12}$ of material that fell in a given time interval, 
$t_1 - t_2$ after $t_0$,
relative to the amount that fell during the full declining part
of the impact curve, $t=t_0$ to $t \rightarrow \infty$, 
can be calculated from Eq.\,\ref{n(t)} as
\begin{eqnarray}
\label{nt1t2}
F_{12} &=& \int_{t_1}^{t_2}e^{-(t-t_0)ln2/\tau_{1/2}}dt \ / 
            \int_{0}^{\infty}e^{-tln2/\tau_{1/2}}dt
                                                        \nonumber \\
       &=& \lbrack e^{-ln2(t)/\tau_{1/2}}\rbrack ^{t=t_2}_{t=t_1}\  /
           \ \lbrack e^{-ln2(t)/\tau_{1/2}}]^{t\rightarrow\infty}_{t=0}
                                                        \nonumber \\
       &=& e^{-ln2(t_1-t_0)/\tau_{1/2}} - e^{-ln2(t_2-t_0)/\tau_{1/2}}
\end{eqnarray}
If we let $t_1$ be the time of the terrestrial peak,
and the impactors were comets, then $t_1-t_0 = 0.75 \tau_{1/2}$
in accordance with Eq.\,\ref{lhbdelay}, and
Eq.\,\ref{nt1t2} simplifies to 
\begin{equation}
\label{ntht2}
F_{12} = 0.6 - e^{-ln2(t_2-t_0)/\tau_{1/2}}
\end{equation}

We will now assume that the cosmic material we find in the Isua
sediments is a result of mixing into an Isua proto-crust that 
existed from the 
time of the terrestrial LHB peak until 3.8 Ga when the sediments formed.
Then $t_2$ = 3.8\,Ga in Eq.\,\ref{ntht2} (and \ref{lhbdelay})
and $t_0$ is the lunar LHB peak.
For reasonable choices of $t_0$ and $\tau_{1/2}$ (e.g., 3.9Ga/50Ma,
3.85Ga/30Ma, 3.95Ga/75Ma), Eq.\,\ref{ntht2} then gives us that
$F_{12} \approx 0.3$.
Since the gravitational focusing of comets toward Earth was found
to be 1.3 times the focusing toward the Moon 
($N(t_0)^\oplus/N(t_0)^{moon}$ = 1.3), the total amount of 
cometary mass that fell on the proto-Isua crust was 
$\sim$ 1.3 $\times$ 0.3 = 40\% 
of what fell on the Moon (and gave rise to the craters we 
see today), under the above assumptions. Half of this mass was found 
to mix with the crust (and half to escape the Earth in high-velocity
plumes). Since the lunar crater density showed us that the impacting
mass on the Moon was $\approx$\,10$^{20}$\,kg, or 2.6\,10$^6$\,kg/m$^2$,
the amount of material that will have mixed with the proto-Isua crust
was 0.2 $\times $2.6\,10$^6$ $\approx$ 5\,10$^5$\,kg/m$^2$
(and the total amount that fell onto it was twice this number, giving
the 1000\,t/m$^2$ stated in the abstract and introduction).

In order to estimate which abundance of iridium this amount of cometary
mass would give rise to in the Isua sediments, we will need to know
the abundance of iridium in comets, and we will need to know how deep
the material mixed into the proto-Isua crust, both of which can at
present unfortunately only be a rough estimate. 
The concept ``comets" represent a wide class of objects,
that are likely to have widely
different compositions. However a reasonable estimate of a representative
``standard cometary composition" could be
80\% ice, 10\% CI material, and 10\% other kinds of dust 
(Festou et al.\ 1993).
If this material mixed homogeneously with the $\approx$ 50 km upper layer 
of the Earth, we reach a fraction of 
(5\,10$^4$\,g/cm$^2$)\,/\,(5\,10$^6$\,cm)\,/\,(3.5\,g/cm$^3$)\,/\,10 =
2.86\,10$^{-4}$\,g CI-material per g crust 
(assuming $\rho_{\rm crust}$\,=\,3.5\,g/cm$^3$). 
With an abundance of 465,000 ppt Ir in CI, 
this finally leads to an estimated 
2.86\,10$^{-4}$\,$\times$\,4.65\,10$^{5}$ $\approx$\,130 ppt 
iridium concentration 
in the Isua proto-crust; a value which is in close agreement
with our measured average concentration of $\approx$\,150 ppt. 

\begin{table}
{\small
\begin{tabular}{|lr|}
\hline
\hline
  Identification   &  Concentration in ppt  \\
\hline
 462906                 &  600$\pm$50     \\
 462948                 &  120$\pm$40     \\
 463418                 &  210$\pm$50     \\
 463428                 &  290$\pm$40     \\
 2000 4                 &  140$\pm$40     \\
 2000 6                 &  $<$ 50         \\
\hline
 simple average         &  230$\pm$50     \\
 average excl.\ sample  &                 \\
 462906 and 2000 6      &  190$\pm$40     \\
\hline
\end{tabular}
\caption{Measured iridium concentration in ppt in six samples 
of Isua metabasalt rocks.}
\label{table2}
}     
\end{table}
Our assumed mixing depth of $\approx$\,50\,km 
seems likely in the light of the high impact kinetic energy,
and is in good agreement with
what has been argued for by others (e.g.\ Sleep et al.\ 1989 argues
for a 35 km mixing depth on the Moon), but even $\approx$ 25 or 100 km mixing
would obviously be in agreement with our measurements (predicting
$\approx$ 260 respectively 65 ppt cometary iridium, still in good 
agreement with our measured 150 ppt within the theoretical uncertainty).
In order to test our assumption about a relatively deep mixing,
we also measured the iridium abundance in 6 samples of metabasaltic 
rocks collected at Isua. The results of these measurements are 
contained in Table 2 and show elevated iridium concentrations in 
these metabasalts, averaging 230$\pm$50 ppt. 
This implies that impact-derived iridium, potentially deposited
on a mafic proto-crust, was transported down into to the mantle where it
was imparted to the melts that later on produced the boninitic and
picritic basalts now exposed in the Isua greenstone belt. Such a
recycling scenario is compatible with the intra-oceanic geotectonic
setting (opening of an asthenospheric window developed as a consequence
of ridge subduction beneath an early Archean arc-forearc region)
proposed by Polat \&\ Frei (2005) to explain the geochemical features of
these metabasalts. 

If the LHB impactors had been asteroids instead of comets, the 
focusing factor would have been 1.8 instead of 1.3, the fraction that 
had mixed into the proto-Isua crust would have been 0.9 instead of 
0.5, and the amount of CI material relative the amount of impacting
material would have been 1 instead of 0.1. Under the same assumptions
used for the estimates regarding cometary impactors above, this 
would therefore have lead to an estimated iridium abundance of
(1.8/1.3)$\times$(0.9/0.5)$\times$(1.0/0.1)$\times$130\,ppt =
3,200\,ppt which obviously is much further from our measured 150\,ppt
than the cometary 130\,ppt. Assuming 35 km mixing depth on the 
Moon (as in Sleep et al.\ 1989) 
and 50 km on the Earth, the predicted iridium abundance in the lunar 
soil due to an asteroidal LHB would be 
3,2000\,ppt/1.8$\times$(0.5/0.9)$\times$(50/35) =
1,400\,ppt. Even though this is still smaller than the predicted terrestrial
iridium abundance for an asteroidal LHB, 
it is obviously in severe disagreement with the
very low measured lunar value of $<$10\,ppt. 

As for the Earth, a 
cometary LHB is, however, in quantitative agreement also with the 
lunar iridium value, because the calculations above showed the 
amount of impacting material to accumulate on the Moon
during a cometary LHB would be (close to) zero, in good 
agreement with the very low lunar value measured.
Accepting a longer mixing history than from the LHB peak and onward,
would not affect the expected lunar Ir abundance if caused by
comets, but would increase it further beyond an already 
unrealistically high value if the impactors were asteroids.

We therefore see that only comets would be able to explain the
profound difference in lunar and terrestrial iridium abundances,
and that in addition reasonable estimates for the period of LHB
mixing, the crustal mixing depth, etc, can lead to a 
quantification in good agreement with the actually measured iridium 
abundance both on Earth and on the Moon. \\

\noindent
{\large \bf Implications for extraterrestrial delivery of water to 
the Earth.}

There exist an extensive literature on the question of whether 
the Earth's oceans (and its atmosphere) originated from geological out-gassing
or from late extraterrestrial impacts by volatiles (and if so, then
by which type of objects). The debate has 
been summarised recently by e.g.\ Delsemme (2006).
It is beyond the goal of the present paper to enter into
this discussion, but our results have two novel inputs to the 
debate that we will explain in this section: 
(1) in order to explain our measured iridium abundances,
an amount of water corresponding to 
a substantial fraction (or all) of the  water in the Earth's present-day 
oceans will have been delivered to the Earth during the LHB
in the form of cometary ice; 
(2) the simple formulas derived for the ratio between mixed and re-emitted
material as function of impactor type, may indicate the existence of 
a feed-back mechanism that in a simple way explains why the combined
size of the Earth's oceans is approximately the size it is.

None of the  theories that have been proposed (neither out-gassing nor 
delivery by any single type of cosmic object) is in agreement with the bulk
of all the isotopic and elemental ratios in the ocean and the atmosphere. 
However, it has often been seen as a strong argument against a cometary origin
of the oceans that three comets (Halley, Hale-Bopp, and Hayakutake) all
have been measured to have a D/H ratio twice the value of the ocean water.
Several papers have summarised the D/H problematics; e.g.\
Robert et al.\ (2000), Morbidelli et al.\ (2000), Delsemme (1999, 2006).
We will therefore here only very shortly 
summarise the relevant numbers for D/H before
proceeding to the explanation of the implications of our Isua iridium
measurements for the delivery of cometary water to the Earth during the LHB.

The D/H ratio in the present-day standard mean ocean water (SMOW) is 156 ppm
(Lodders \& Fegley 1998), the D/H ratio in the proto-solar nebula 
(out of which the Earth formed) is inferred to have been 25 ppm from
observations of the present-day D/H ratio in the atmosphere of Jupiter
and Saturn (26$\pm$7 ppm and 25$\pm$10 ppm for respectively Jupiter and
Saturn; Lodders \& Fegley 1998), 
and the proto-ices (i.e., comets) in the Neptune region is inferred
to have been between 70 ppm and 250 ppm (Lecluse et al.\ 1996,
Mousis et al.\ 2000) based on models and measured values (120 ppm;
Lodders \& Fegley 1998) of Neptune's atmosphere. 
The water on Mars is inferred from SNC
meteorites to have a D/H ratio of 300 ppm (Leshin 2000).
Most CI, CM, and CV meteorites have values between 130 and 170 ppm
(i.e., close to SMOW), CR meteorites and ordinary chondrites have values
around 250 ppm (with large variations) (Robert 2003).
The three measured values of comets (Halley, Hale-Bopp, and Hayakutake)
was found to $\sim$300 ppm (e.g.\ Bockel{\'e}e-Morvan 1998).
Based on these numbers, a "cometary value of D/H" has been
assigned to be twice the SMOW value, and is often seen as
an evidence that the oceans of the Earth cannot have had a cometary origin
but must be either from carbonaceous chondrites or intrinsic.
However, as we already pointed out in the beginning of this paper
(in relation to which PGE abundance ratios to expect from impacts)
"comets" is not a homogeneous type of objects, and it is not obvious
which type of comets may have contributed to the LHB. Delsemme (1999)
estimated that with a reasonable mix of comets from different regions
of the outer solar system, the measured D/H in ocean water could be
explained by cometary impacts, and he found that the statistical probability
of yet having measured the class of comets that were most likely to have
contributed to the delivery of the Earth's oceans was quite small.

We argued above that our iridium measurements 
at Isua was in agreement with an estimated total of 10$^{20}$\,kg of 
cometary material having impacted the Moon since the time of the 
(lunar) LHB peak, and that this corresponded to 
2.6\,10$^6$ $\times$ 1.3 kg/m$^2$ of cometary impacts on Earth. 
If water could be treated the same way as our estimates of the
iridium enrichment (i.e., very iridium-rich impactor material hitting
a very iridium-poor target), then half would have remained in the 
Earth's crust and half would have been re-emitted to space in high-velocity 
plumes. Hence, 1700\,t (2600\,$\times$\,1.3/2\,=\,1700) 
of cometary material would have mixed with each
m$^2$ of Earth's dry crust. If comets are 80\% water, as assumed in the 
calculations above, this would correspond to the equivalent of a
2 km deep present-day-area ocean (1.7\,$\times$\,0.8/0.7\,=\,2).

However, the Earth will obviously not stay dry
(i.e., a water-poor target) forever once it is 
impacted with km-layers of ice. Interestingly, this causes a feedback
mechanism that prevents the Earth from being completely transformed into 
an ocean planet, independently of how many comets impacted the 
Earth during the LHB. This is because the ratio between delivery and 
expelling of water during impact depends on whether comets (of the 
assumed velocity distribution) hits ocean or dry land. In the simple 
theory outlined above, we argued that it happens to be so that 
the meridian velocity of Earth crossing comets is roughly equal to 
the velocity required to form a high velocity plume
(Eq.\,\ref{vminplume}$-$\ref{vminval}). When such a plume
forms, all the cometary mass escapes back into space, together with
a similar amount of target mass. On top of this, a typical impact
accelerates additional (surrounding) target mass equivalent 
to $\sim$20\% of its mass into escape velocity 
(Eq.\,\ref{mejcti2} with $\rho_t \approx \rho \approx 1$\,g/cm$^3$
and $v \approx 2 v_{esc}$).
For comets hitting an ocean, the target mass is water, and therefore
50\% of the LHB comets that hit the ocean will not only disappear 
back into space in the plume, but they will drag a similar amount
of ocean water into space, plus a minor amount of surrounding target
mass (usually also ocean-water). In summary, the impact in the ocean
of a high velocity (i.e., $v > v_{esc}$ in Eq.\,\ref{mejcti2})
comet of mass $m$ will cause the Earth to lose  
$\sim$\,1.2$m$ of water. A similar comet of too low velocity to cause
an Earth-escaping plume, will deliver $(0.8\,-\,0.2)\,m\,=\,0.6\,m$ worth 
of water. For comets hitting dry land, the corresponding budget is
zero water delivery for the high velocity comets, and 
$0.8\,m$ for the low-velocity ones. When most of the Earth is still
dry land, the delivery budget is therefore that 40\% of the impacting cometary
mass will stay in the form of delivered water. Once the created ocean
covers $\sim$50\% of the Earth's surface, the budget will have 
become: The ocean hitting comets subtract $\sim\,(0.6/2-1.2/2)\,m = -0.3\,m$
of water, and the 50\% land hitting comets will 
add $\sim\,(0.4/2)\,m = 0.2\,m$ water.
Hence, when the Earth's ocean-covered increases beyond 50\%, more
water will be subtracted than added on average, until the cover again
decreases below $\sim 50\%$. This feed back mechanism
would therefore result in that the Earth is covered with oceans on 
$\sim$\,half of its surface, in rough agreement (within the 
estimates given) with the actual present-day ocean cover.
In summary, this simple estimate
would reduce the delivered present-day-area ocean from the 2 km depth 
calculated above to, say, $\sim$1 km depth,
but still, obviously, being a considerable amount. 

Chyba (1990) computed the effect of cometary impacts in a
static ocean (and therefore obviously didn't reach conclusions about
a possible maximum extend). 
Under the assumption of a more restricted interaction
between the ocean and the impactor (including that no direct ejecta
is created and that only water initially in front of the impactor
on its passage through the ocean will escape the Earth), he
reached the conclusion that only up to 15\% of the ocean could be eroded
away because of high-velocity cometary impacts. A more thorough analysis,
beyond the scope of the present work, would have to take into 
account the statistics of the distribution of the few highest mass 
impacts, the effect of ocean-evaporating impacts, the loss of 
water with atmosphere-stripping impacts, a more precise 
estimate of the abundance of water in the LHB comets, an estimate of the 
topography of the early Earth, etc. However, it is clear from
the very simple estimates above that there may be a natural 
upper limit of the size of an impact delivered ocean, and that 
the water associated with the iridium enrichment we have measured at Isua
will have been a substantial fraction of the present-day ocean mass,
if our interpretation, that the iridium in the 3.8 Ga old Isua rocks
represents a cometary LHB impact, is correct.

In this way, our measurements are the 
first results to bring a conceptual and quantitative agreement between the 
lunar LHB cratering records and the geochemical records on the part of the
Earth's crust that existed during the LHB period.
The results are in good qualitative agreement with recent models 
(Gomes et al.\ 2005)
for the solar system formation, 
which predict a large fraction of the LHB to be 
comets perturbed onto collision course with the Earth and Moon when Jupiter and 
Saturn migrated through a 1:2 orbital resonance 3.9 Ga ago, 
but which also predict
that a large fraction could be asteroids (in contradiction with our
measurements). Our results would be in disagreement with
the expected Isua proto-crustal (and lunar) Ir abundance if the LHB impactors 
were asteroids,
unless some unknown mechanism would be able to remove the iridium from the 
solid crust (on Earth as well as on the Moon) in 
which the asteroids impacted.\\

\newpage

\noindent
{\large \bf Conclusions.}

We have sampled 3 different types of sedimentary rocks from the 
Isua greenstone belt in Greenland, which with an age of 
$\sim$3.8\,Ga is the oldest known major piece of the Earth's crust.
We argued that the 3 types of metasediments
potentially contain an average proto-crustal signature transported by
rivers to the site of deposition (in case of clastic sediments) and
likely reflect a contemporaneous dissolved seawater inventory in case of
the chemical metasediments (BIFs). 37 individual samples were 
mortared and neutron-radiated in a nuclear research reactor.
Subsequent $\gamma-\gamma$ coincidence spectroscopy of the
radiated samples revealed an average iridium abundance of $\approx$150 ppt.
This is an enhancement of the
Isua proto-crust relative to the Earth's present-day ocean and
upper crust of a factor 7, and relative
to the lunar crust with a factor of more than 15.

We argued that this enrichment is in 
qualitative agreement with a cometary LHB, but in qualitative 
disagreement with a LHB caused by asteroids.

In order to also see which quantitative restrictions the measured
iridium abundances could impose on various impact scenarios, we
developed the theoretical basis for how the lunar crater
counts can be scaled to LHB impact mass on the Isua proto-crust.
We first toughened the concept of LHB peak up a bit and quantified 
the delay in the terrestrial LHB peak relative to the 
lunar LHB peak (usually just called the LHB peak), and explained
how the shift in the time of the peak, as well as the whole
terrestrial LHB energy density flux curve,
depends on whether the impactors were comets or asteroids.
We then quantified how impacts caused by asteroids differ from
impacts caused by comets, in terms of the amount of iridium they
will have imposed into the Isua proto-crust and into the lunar surface. 
By use of the developed framework and selection of reasonable
values for the relevant parameters (such as mixing depth, 
mixing duration, etc), we estimated under which conditions a 
cometary LHB on Earth would give rise to roughly
the measured iridium abundance in the sampled Isua sediments,
and at the same time also to the measured value in the lunar surface 
material. As far as we know, it is the first time that a 
self-consistent scenario has been presented in the literature,
which rigorously quantifies the scaling of a cometary lunar LHB to the 
conditions on the Earth, and is able to find agreement between
the lunar cratering counts and elemental abundances measured
in the lunar soil as well as in the early terrestrial crust.

A similar calculation based on the assumption that the LHB impactors
were asteroids (i.e., ''meteorites") lead to a predicted 
Isua iridium abundance approximately a factor 20 higher than we 
measured, and a factor of several hundreds too high for the lunar 
abundance. 
Therefore comets but not asteroids can quantitatively account for the 
measured values of both the lunar and Isua iridium abundances,
as well as the impacting mass that have given rise to 
the lunar craters. 

The quantification made it possible
for us to estimate the total (cometary) mass that have 
hit the Earth, and the (impactor-dependent) balance between 
impacting and re-emitted material.
We therefore finally estimated what effect a cometary LHB impact can 
have had on the formation of the Earth's oceans. We discovered that 
there is a feed back mechanism that will prevent the oceans caused
by cometary impacts to cover more than approximately 50\% of the
Earth's surface, and calculated that the LHB cometary impactors 
that can have caused the craters we see on the Moon today and 
explain the iridium abundances measured in the Isua sediments and 
in the lunar soil, will have covered the Earth with a km deep ocean
over $\sim$50\% of its surface.\\

\noindent
{\large \em Acknowledgements:}\\
Valuable comments from S.Moorbath, A.Polat, B.Reipurth,
R.Gwozdz, and the two referees, are greatly acknowledged. 
This work was supported 
by the Danish Natural Science Research Council (FNU) and 
the Isua Multidisciplinary Research Project (IMRP).

\newpage
\noindent
{\large \bf References:}

\vspace{5mm}

{\small
\begin{list}{}{\parsep=0mm\itemsep=2.5mm\leftmargin=0mm}

\item[]
Alexander,B.W. Bau,M., Andersson,P., Dulskiet,P. 2008.
Continentally-derived solutes in shallow
Archean seawater: Rare earth element and Nd isotope evidence in iron
formation from the 2.9 Ga Pongola Supergroup, South Africa.
Geochimica et Cosmochimica Acta 72, 378$-$394.

\item[]
Alvarez,L.W. W.Alvarez,W., F.Asaro,F., H.V.Michel,H.V., 
1981. Asteroid Extinction Hypothesis. Science 211, 654$-$656.

\item[]
Anbar,A.D., Zahnle,K.J., Arnold,G.L.,  Mojzsis,S.J. 2001.
Extraterrestrial iridium, sediment accumulation and the habitability of
the early Earth's surface. J.\ Geophys.\ Res.\ 106, 3219$-$3236.

\item[]
Appel,P.W.U., Fedo,C.M., Moorbath,S., Myers,J.S., 1998. 
Recognizable primary volcanic and sedimentary features in a low-strain 
domain of the highly deformed, oldest known (~3.7$-$3.8 Gyr) 
Greenstone Belt, Isua, West Greenland. Terra Nova 10, 57$-$62.

\item[]
Bockel{\'e}e-Morvan,D., Gautier,D., Lis,D.C., Young,K., Keene,J.,
Phillips,T., Owen,T., Crovisier,J., Goldsmith,P.F., Bergin,E.A.,
Despois,D., Wooten,A. 1998.
Deuterated water in comet  C/1996 B2 (Hyakutake) and its implications
for the origin of comets.  Icarus 193, 147$-$162.

\item[]
Bohor,B.F., Modreski,P.J., Ford,E.E., 1987. 
Shocked quartz in the cretaceous-tertiary boundary clays -- 
Evidence for a global distribution.
Science 236, 705$-$710.

\item[]
Bolhar,R., Kamber,B.S., Moorbath,S., Fedo,C.M., Whitehouse,M.J., 2004.
Characterisation of early Archaean chemical sediments by trace element 
signatures. Earth Planet.\ Sci.\ Let.\ 222, 43$-$60.

\item[]
Bolhar,R., Kamber,B.S., Moorbath,S., Whitehouse,M.J., Collerson,K.D.,
2005. Chemical characterization of Earth's most ancient metasediments
from the Isua Greenstone Belt, southern West Greenland.
Geochimica et Cosmochimica Acta 69, 1555$-$1573.

\item[]
Bronshten,V. 2000. 
On the nature of the Tunguska meteorite.
Astron.\ Astrophys.\ 359, 777$-$779.

\item[]
Chyba,C.F., 1990. Impact delivery and erosion of planetary
oceans in the early inner Solar System.
Nature 343, 129$-$133.

\item[]
Chyba,C.F., 1991. Terrestrial mantle siderophiles and the 
lunar impact record. Icarus 92, 217$-$233.

\item[]
Cisowski,S.M. 1990. 
A critical review of the case for, and against, 
extraterrestrial impact at the K/T boundary.
Surveys in Geophysics 11, 55$-$131.

\item[]
Dauphas, N., 2003. The dual origin of the terrestrial atmosphere. 
Icarus 165, 326-339.

\item[]
Delsemme, A. 1999. The deuterium enrichment in recent comets is
consistent with the cometary origin of seawater.
Planetary and Space Sci.\ 47, 125$-$131.

\item[]
Delsemme, A. 2006. The origin of the atmosphere and the oceans.
In: P.J.Thomas, R.D.Hicks, C.F.Chyba, C.P.McKay (eds.), Comets and the
origin and evolution of life. Springer, 29$-$68.

\item[]
Fedo,C.M., Whitehouse,M.J., 2002.
Metasomatic origin of quartz-pyroxene rock, Akilia, Greenland,
and implications for Earth's earliest life.
Science 296, 1448$-$1452.

\item[]
Festou,M.C., Rickman,H., West,R.M., 1993.
Comets. I - Concepts and observations.
Astron.\ Astrophys.\ Rev.\ 4, 363$-$447.

\item[]
Frei,R. Frei,K.M., 2002. A multi-isotopic and trace element investigation 
of the Cretaceous-Tertiary boundary layer at Stevns Klint, Denmark
-- inferences for the origin and nature of siderophile and lithophile 
element geochemical anomalies. Earth Planet.Sci.Let.\ 203, 691$-$708.

\item[]
Frei,R., Polat,A., 2007. Source heterogeneity for the major components of
$\sim$3.7 Ga Banded Iron Formations (Isua Greenstone Belt,
Western Greenland): Tracing the nature of interacting water masses in BIF 
formation. Earth Planet.\ Sci.\ Let. 253, 266$-$281.

\item[]
Frei,R., Rosing,M.T., 2005. 
Search for traces of the late heavy bombardment on Earth --
results from high precision chromium isotopes.
Earth Planet.\ Sci.\ Let.\ 236, 28$-$40.

\item[]
Glikson, A.Y. 2008.
Field evidence of Eros-scale asteroids and impact-forcing
of Precambrian geodynamic episodes, Kaapvaal (South Africa) and Pilbara
(Western Australia) Cratons. 
Earth Planet.\ Sci.\ Lett.\  267, 558$-$570.

\item[]
Glikson, A.Y. 2004. 
An Alternative Earth: Comment. 
GSA Today, 10.1130/1052-5173(2004)014<e1, e1-e2.

\item[]
Gomes,R., Levison,H.F., Tsiganis,K., Morbidelli,A., 2005.
Origin of the cataclysmic Late Heavy Bombardment period of the 
terrestrial planets.
Nature 435, 466$-$469.

\item[]
Gros,J., Takahashi,H., Hertogen,J., Morgan,J.W., Anders,E., 1976.
Composition of the projectiles that bombarded the 
lunar higlands. Proc.\ Lunar Sci.\ Conf.\ 7, 2403$-$2425.

\item[]
Hansen,H.J., Gwozdz,R., Rasmussen,K.L., 1988.
High-resolution trace element chemistry across the Cretaceous-Tertiaty
boundary in Denmark. Rev.Esp.Peleontol., 21-29.

\item[]
Hartmann,W.K., Ryder,G., Dones,L., Grinspoon,D., 2000.
The time dependent intense bombardment of the promordal
Earth/Moon system.
 In Origin of the Earth and Moon (eds.\ R.Canup, K.Righter) 
(Univ.\ Arizona Press, Tucson), 493$-$512.

\item[]
Hatsukawa,Y., Mahmudy Gharaie,M.H., Matsumoto,R., Toh,Y.,
  Oshima,M., Kimura,A., Noguchi,T., Goto,K., Kakuwa,Y., 2003.
Ir anomalies in marine sediments: case study for the Late Devonian 
mass extinction event. Geochim.\ Cosmochim.\ Acta Suppl.\ 67, 138$-$147.

\item[]
Hatsukawa,Y., Miyamoto,Y., Toh,Y., Oshima,M., 
  Hosein Mahmudy Gharaie,M., Goto,K., Toyoda,K., 2007.
  J.\ Radioanalyt.\ Nucl.\ Chem.\ 272, 273.

\item[]
Hatsukawa,Y., Oshima,M., Hayakawa,T., Toh,Y., Shinohara,N., 2002.
Application of multiparameter coincidence spectrometry
using a Ge detectors array to neutron activation analysis.
 Nucl.\ Instrum.\ Methods Phys.\ Res.\ A482, 328$-$333.

\item[]
Hertogen,J., Janssens,M.-J-, Takahashi,H., Palme,H., Anders,E., 1977.
Lunar basins and craters: Evidence for systematic compositional 
changes of bombarding population.
Proc.\ Lunar Sci.\ Conf.\ 8, 17$-$45.

\item[]
Hong,Z., Weiguang,Z., Liang,Q., Meifu,Z., Yieyan,S., Yi,Z.
2006. Platinum-group element (PGE) geochemistry of the Emeishan basalts
in the Pan-Xi area, SW China.

\item[]
Hou,Q.L. P.X.Ma,P.X. E.M.Kolesnikov,E.M., 1998.
Discovery of iridium and other element anomalies near the 1908 Tunguska
explosion site, Planet.\ Space Sci.\ 46, 179$-$188.

\item[]
Jacobsen,S.B., 2005.
The Hf-W isotopic system and the origin of the Earth and Moon.
Ann.\ Rev.\ Earth Planet.\ Sci.\ 33, 531$-$570.

\item[]
Jenner,F.E., Bennett,V.C., Nutman,A.P., Friend,C.R.L., Norman,M.D.,
Yaxley,G., 2008.  
Evidence for subduction at 3.8 Ga: Geochemistry of arc-like metabasalts
from the southern edge of the Supracrustal belt.
Chemical Geology, in press.

\item[]
Jopek,T.J., Froeschl{\'e},C., Gonczi,R., Dybczy{\'n}ski,P.A. 2008. 
Searching for the parent of the Tunguska cosmic body.
Earth Moon Planet 102, 53$-$58.

\item[]
Kamber, B.S., Moorbath, S. 1998.
Initial Pb of the Amitsoq gneiss revisit: implications for the 
timing of early Archean crustal evolution in West Greenland.
Chem.\ Geo.\ 150, 19$-$41.

\item[]
Kamber, B.S., Moorbath, S. 2000.
Initial Pb of the Amitsoq gneiss revisit: implications for the 
timing of early Archean crustal evolution in West Greenland $-$ Reply.
Chem.\ Geo.\ 166, 309$-$312.

\item[]
Kamber, B.S., Whitehouse, M.J., Bolhar, R. 2005.
Volcanic resurfacing and the early terrestrial crust: Zircon U-Pb and
REE constraints from the Isua Greenstone Belt, southern West Greenland.
Earth Planet.\ Sci.\ Let.\ 240, 276$-$290.

\item[]
Koeberl,C., Reimold,W.U., McDonald,I., Rosing,M., 1999. 
Search for petrographic and geochemical evidence for the 
late heavy bombardment on Earth in early Archean rocks from
Isua Greenland.
in: I.Gilmour\& C.Koeberl (eds.): Impacts and the Early Earth, 
  (Springer-Verlag, Berlin), 73$-$97.

\item[]
Kolesnikov,E.M., Hou,Q.L., Xie,L.W., Kolesnikova,N.V., 2005.
Finding of probable Tunguska cosmic body material: anomalies in platinum
group elements in peat from the explosion area.
  Astronomical and Astrophysical Transactions 24, 101-111.

\item[]
Korina,M.I.. Nazarov,M.A., Barsukova,L.D., Suponeva,I.V., Kolesov,G.M.,
  Kolesnikov,E.M., 1987.
Iridium distribution in in the peat layers from area of Tunguska event.
 Lunar Planet.\ Sci.\ Conf.\ 18, 501.

\item[]
Kring,D.A., Cohen,B.A., 2002.
Cataclysmic bombardment throughout the inner solar system 3.9-4.0 Ga.
J.Geophys.Res.\ 107, E2, 4-1$-$4-6.

\item[]
Lecluse,C., Robert,F., Gautier,D. Guiraud,M. 1996.
Deuterium enrichment in giant planets. 
Planet.\ Space Sci.\ 44, 1579$-$1592.

\item[]
Leshin,L.A. 2000.
Insights into Martian water reservoirs from analyses of Martian meteorite
QUE94201.
Geophys.\ Res.\ Lett.\ 27, 2017$-$2020.

\item[]
Lodders,K., Fegley,B.Jr., 1998. The Planetary Scientist's Companion.
Oxford Univ.\ Press, Oxford.

\item[]
Lowe,D.R., Byerly,G.R., Kyte,F.T., Shukolyukov,A., Asaro,F., 
Krull,A. 2003. 
Spherule beds 3.47-3.34 Ga-old in the Barberton Greenstone Belt, South
Africa: a record of large meteorite impacts and their influence on early
crustal and biological evolution. 
Astrobiology 1, 7$-$48.

\item[]
Maier, W.D., Roelofse, F., Barnes, S.-J. 2003. The concentration
of the platinum-group elements in South African komatiites: implications
for mantle sources, melting regime and PGE fractionation during
crystallization. Journal of Petrology, 44, 1787$-$1804.

\item[]
Martin,H., Albar{\`e}de,F., Claeys,P., Gargaud,M., Marty,B.,
Morbidelli,A., Pinti,D.L., 2006. Building of a habitable planet.
Earth, Moon, and Planets 98, 97$-$151.

\item[]
Marty,B., Meibom,A., 2007.
Noble gas signature of the late heavy bombardment in the 
Earth's atmosphere. eEarth 2, 43$-$49.

\item[]
Melosh, H.J., 1989. Impact Cratering -- a geological process.
Oxford Monographs Geol.\ Geophys.\ 11, Oxford Univ.\ Press.

\item[]
Melosh, H.J., Vickery, A.M., 1989. Impact erosion of the primordial 
atmosphere of Mars. Nature 338, 487-489.

\item[]
Mojzsis, S.J., Harrison, T.M. 2000.
Vestiges of a beginning: Clues to the emergent biosphere in the
oldest known sedimentary rocks. GSA Today 10, 1$-$6.

\item[]
Moorbath,S. 2005. Dating earliest life. Nature 434, 155.

\item[]
Morbidelli,A., Chambers,J., Lunine,J.I., Petit,J.M., Robert,F.,
Valsecchi,G.B., Cyr,K.E.,
2000. Source regions and time scales for the delivery of water to Earth. 
Meteoritics \&\ Planetary Science 35, 1309$-$1320.

\item[]
Mousis,O., Gautier,D., Bockel{\'e}e-Morvan, D., Robert, F., 
Dubrulle,B. Drouart,A. 2000.
Constraints on the formation of comets from D/H ratios measured in
H$_2$O and HCN. Icarus 148, 513$-$525.

\item[]
Nutman, A.P., McGregor, V.R., Friend, C.R.L., Bennet, V.C., 
Kinny, P.D. 1996. The Itsaq Gneiss Complex of southern West Greenland: 
the worlds most extensive record of early crustal evolution 
(3900$-$3600 Ma). Precambrian Research 78, 1$-$39.

\item[]
Nutman, A.P., Mojzsis, S.J., Friend, C.R.L. 1997.
Recognition of $>$3850 Ma water-lain sediments in West Greenland
and their significance for the early Archean Earth.
Geochim.\ et.\ Cosmochim.\ Acta 61, 2474$-$2484.

\item[]
Oshima,M., Toh,Y., Hatsukawa,Y., Hayakawa,T., Shinohara,N., 2002.
 J.\ Nucl.\ Sci.\ Technol.\ 39 292.

\item[]
Oshima,M., Toh,Y., Hatsukawa,Y., Koizumi,M., Kimura,A., Haraga,A.
 Ebihara,M., Sushida,K., 2008.
J.\ Radioanal.\ Nucl.\ Chem.\ submitted.

\item[]
Owen,T., Bar-Nun,A., Kleinfeld,I., 1992.
Possible cometary origin of heavy noble gases in the atmospheres 
of Venus, earth, and Mars. Nature 358, 43$-$46.

\item[]
Owen,T., Bar-Nun,A., 2001.
Contributions of Icy Planetesimals to the Earth's Early Atmosphere.
Origins of Life and Evolution of the Biosphere 31, 435$-$458.

\item[]
Pater, I.de, Lissauer, J.J. 2001.
Planetary Sciences, Camb.Univ.Press., Cambridge.

\item[]
Pepin, R.O., 1991. On the origin and early evolution of terrestrial 
planet atmospheres and meteoritic volatiles. Icarus 92, 2-79.

\item[]
Polat, A., Frei, R., 2005. The origin of early Archean banded iron 
formations and of continental crust, Isua southern West Greenland. 
Precambrian Research 138, 151$-$175.

\item[]
Polat,A., Hofmann,A.W., Muenker,C., Regelous,M., Appel,P.W.U. 2003.
Contrasting geochemical patterns in the 3.7$-$3.8 Ga pillow basalt cores
and rims, Isua Greenstone belt, Southwest
Greenland; implications for postmagmatic alteration processes. Geochim.
Cosmochim.\ Acta 67. 441$-$457.

\item[]
Polat, A., Hofmann, A.W., Rosing, M.T., 2002. Boninite-like volcanic rocks in 
the 3.7-3.8 Ga Isua greenstone belt, West Greenland: geochemical evidence 
for intra-oceanic subduction zone processes in the early Earth. 
Chemical Geology 184, 231$-$254.

\item[]
Puchtel, I., Humayun, M. 2000. Platinum group elements in the
Kostomuksha komatiites and basalts: implications for oceanic crust
recycling and core-mantle interaction. 
Geochimica et Cosmochimica Acta 64, 4227$-$4242.

\item[]
Rasmussen,K.L. Clausen,H.B. Lallemeyn,G.W., 1995. No iridium after the 1908 
Tunguska impact: Evidence from a Greenland ice core. 
Meterortics 30, 634$-$638.

\item[]
Rasmussen,K.L. Olsen,H.J.F. Gwozdz,R. Kolesnikov,E.M., 1999. 
Evidence for a very high carbon/iridium ratio in the Tunguska
impactor. Meteoritics \&\ Planetary Science 34, 891$-$895.

\item[]
Robert,F. 2003.
The D/H ratio in chondrites.
Space Science Reviews 106, 87$-$101.

\item[]
Robert,F., Gautier,D., Dubrulle,B., 2000.
The Solar System D/H Ratio: Observations and Theories.
Space Science Reviews 92, 201$-$224.

\item[]
Rosing,M.T., 1999.
$^{13}$C-depleted carbon microparticles in $>$3700-Ma sea-floor sedimentary
rocks from West Greenland. Science 283, 674$-$676.

\item[]
Ryder,G., 2002. Mass flux in the ancient Earth-Moon system and 
benign implications for the origin of life on Earth.
 J.\ Geophys.\ Res.\ 107, E4, 6-1:6-13.

\item[]
Schoenberg,R., Kamber,B.S., Collerson,K.D., Moorbath,S., 2002.
Tungsten isotope evidence from $\sim$3.8-Gyr metamorphosed sediments 
for early meteorite bombardment of the Earth.
Nature 418, 403$-$405.

\item[]
Shukolyukov,A., Lugmair,G.W., 1998.
Isotopic evidence for the Cretaceous-Tertiary impactor and its type.
Science 282, 927$-$929.

\item[]
Simonson,B.M., Glass,B.P. 2004. 
Spherule layers $-$ records of ancient impacts. 
Ann.\ Rev.\ Earth Planet.\ Sci.\ 32, 329$-$361.

\item[]
Sleep,N.H., Zahnle,K.J., Kasting,J.F., Morowitz,H.J., 1989.
Annihilation of ecosystems by large asteroid impacts on the early Earth.
Nature 342, 139$-$142.

\item[]
Snyder,G.A., Borg,L.E., Nyquist,L.E., Taylor,L.A., 2000.
Chronology and isotopic constraints on lunar evolution.
 In Origin of the Earth and Moon (eds.\ R.Canup, K.Righter) 
(Univ.\ Arizona Press, Tucson), 361$-$396.

\item[]
Toh,Y., Oshima,M., Hatsukawa,Y., Hayakawa,T., Shinohara,N., 2001.
Comparison method for neutron activation analysis with $\gamma -\gamma$ 
activation.  J.\ Radioanal.\ Nucl.\ Chem.\ 250, 373$-$376.

\item[]
Turco,R.P., Toon,O.B., Park,C., White,R.C., Pollack,J.B., Noerdlinger,P., 
1982. An analysis of the physical, chemical, optical, and historical impacts 
of the 1908 Tunguska meteor fall.  Icarus 50, 1$-$52.

\item[]
N.V.Vasilyev, N.V., 1998.  The Tunguska Meteorite problem today.
Planet.\ Space Sci.\ 46, 129$-$150.

\item[]
Wedepohl,K.H., 1969. Handbook of Geochemistry. (Springer-Verlag, Berlin).

\item[]
Zahnle,K.J., Sleep,N.H., 1997. 
Impacts and the early evolution of life.
in P.J.Thomas et al (ed.), Comets and 
the origin and evolution of life, 2${\rm nd}$ edition, Springer, 207$-$251.

\item[]
Zahnle,K.J., 2006.
Earth's earliest atmosphere. Elements 2, 217$-$223.

\end{list}
\end{document}